\documentclass{article}
\usepackage[utf8]{inputenc}
\newcommand\x{J}
\usepackage{geometry}
\usepackage{float}
\usepackage{amsmath}
\usepackage{lscape}
\usepackage{rotating}
\usepackage{xcolor}
\usepackage{graphicx}
\usepackage{siunitx}
\usepackage{multirow}
\usepackage{authblk}
\usepackage{wrapfig}
\usepackage{caption}
\usepackage{subcaption}
\usepackage{hyperref}
\usepackage[title]{appendix}
\usepackage{amssymb}
\usepackage[english]{babel}
\usepackage{comment}
\usepackage{csquotes}
\newcommand*{\email}[1]{%
	\normalsize\href{mailto:#1}{#1}\par
}
\title{Study of a Mismatched Parallel Wire Transmission Line and its Potential Applications}

\date{}
\author[1]{Atharva Kulkarni}
\author[1]{Amit Morarka*}

\affil[1]{Department of Electronics and Instrumentation Science, Savitribai Phule Pune University,Ganeshkhind, Pune, Maharashtra, India- 411007\newline \email{*amitmorarka@gmail.com}}

\begin{document}

	\maketitle
	\begin{abstract}
		In this work a mismatched section of a transmission line is examined for its utility in a bandpass filtering application. A model to analyze the situation is given. It was found that the transfer characteristics of such mismatched transmission line resemble with that of a narrow bandpass filter with steeper roll-off.  It has been shown that the center frequency of the response depends on the length of the mismatched section as well as the dielectric material of the line. Further it has been shown that the center frequency can be tuned to a desired value by adjusting the dielectric profile along the length of the line. This was tested on a set of parallel wire transmission lines of three lengths and liquid dielectrics. During the study it was noticed that a sensing mechanism to detect variation in small volumes of the liquid dielectrics with high sensitivity can be developed using mismatched transmission lines.
	\end{abstract}
	
	\section{Introduction}
	\label{Intro}
	In the last couple of decades, re-configurable RF solutions have graduated to include passive as well as active components in order to make designs tunable. Various possibilities have been explored such as using mechanical switches, microelectromechanical devices, monolithic solid state devices, etc. in order to tune to desired frequency. On the other hand, several microfluidic based techniques have been shown to produce similar tunable RF components. These reported techniques (\cite{IP} \cite{Chlieh} \cite{Diedhiou}) involve fabrication of microfluidic channels embedded in microstrip structures. The concept involves manipulating the effective dielectric constant of the substrate in order to get the desired transfer characteristics. Thus a controlled pumping of liquid dielectrics in these channels produces modification of the frequency response of the devices. Such techniques require a sophisticated assembly to control the flow and amount of dielectrics. During the course of present study another area of research came into notice which focuses on measurement of permittivity and permeability of dielectrics using transmission line transducer cells. This involves coaxial transmission line cells which are filled with the dielectric materials. Nicolson-Ross-Weir (NRW) algorithm is used to compute $\mu_r$ and $\epsilon_r$ from the measured S-parameters of the coaxial cells. The mismatch between coaxial cell and the measurement system has been identified as an impediment in accurate computation of $\mu_r$ and $\epsilon_r$. The author in \cite{Leonardo} gives an improvisation to account for the mismatch. A detailed signal flow model along with cell design has been presented in \cite{L.Silva}. 
	
	Till date to the best of our knowledge it is seen that efforts were taken to reduce the mismatch between load and transmission lines \cite{Pozar}\cite{White} in almost all practical applications. There were no attempts to explore the mismatch condition as an opportunity to develop potential applications.	In the present work a novel methodology is examined towards the use of a mismatched transmission line (Tx line) for obtaining re-configurable devices. A bandpass filter is envisaged whose center frequency can be tuned by changing dielectric material of a mismatched Tx line. An attempt of purposely setting the characteristic impedance of a Tx line unequal to the load impedance can be useful and give an advantage of frequency selectivity. In this work the said situation is studied and the results are given. The situation has been analyzed using Tx line theory. It is known from basic Tx line analysis that a mismatched line shows same load impedance after every half a wavelength \cite{Pozar}
	 This property can be used for effective power transmissions only at frequencies at which load impedance does not transform. Such a line shows characteristics of a band pass filter \cite{kuo} \cite{Parraman} with quality factor comparable with other known filter types. The phenomenon discussed in this work is found to be analogues with  the phenomenon of complete transmission of certain wavelengths in a Fabry-Perot Etalon. Like Fabry-Perot Etalon, a section of mismatched Tx line too has filtering capabilities. The phenomenon was modeled and experimentally verified on a set containing parallel wire Tx lines of three distinct lengths with three different liquid dielectrics.

	 Section \ref{theory} establishes the basic principles of the operation along with the model adopted to explain behavior of Tx line. Section \ref{Observations} describes the experimental setup and the method used to dispense liquids on the line. Section \ref{results} gives results of the experiments performed.

	\section{Theory and Background}
	\label{theory}
	A Tx line can be characterized by stating its characteristic impedance and it’s constant of propagation for Electromagnetic (EM) waves. 
	Further both of these depend upon line’s geometry and the materials making it. For any line one can define a set of four parameters viz. Resistance per unit length (R), Conductance per unit length (G), Inductance per unit length (L) and Capacitance per unit length (C) which take their values depending upon both the geometry as well as material. Here parameters R and G together represent losses in the line, whereas L and C represents the energy stored in magnetic field and electric field respectively. For a practical lossless line, both R and G are negligible in comparison with L and C and hence they can be ignored. Eqn(\ref{Char impedance and prop const}) gives the characteristic impedance ($Z_0$) and propagation constant($\gamma = \alpha+j\beta$) for a lossy as well as lossless line \cite{Pozar}\cite{White}  
		\begin{subequations}	
			\begin{align}
				Z_0=\sqrt{\frac{R+j\omega L}{G+j\omega C}}\hspace{0.5cm} \rightarrow \hspace{0.5cm} Z_0=\sqrt{\frac{L}{C}}\\
				\gamma = \alpha+j\beta= \sqrt{(R+j\omega L)(G+j\omega C)}\hspace{0.5cm} \rightarrow \hspace{0.5cm} \gamma = j\beta=j\omega\sqrt{( LC)}\\
			\intertext{It follows from Eqn(\ref{Char impedance and prop const}b) that the phase velocity ($v$) of EM waves is}
			v= \frac{1}{\sqrt{LC}} \propto \frac{1}{\sqrt{\epsilon_r}} \hspace{4cm}
		\end{align}		
		\label{Char impedance and prop const}
		\end{subequations}
	Both L and C depend upon the material between two conductors of Tx line along with the geometry of line. Thus for a fixed geometry, $\epsilon_r$ and $\mu_r$ of dielectric filling the line will decide the characteristics of Tx line. 
	\subsection{Impedance Transformation}
	\label{Impedance Transformation}
	From the solution of the telegrapher’s equation \cite{Pozar}\cite{White}, it is clear that when a traveling incident wave strikes on a mismatched load ($Z_L$),  a reflected wave gets generated in order to preserve the boundary condition at the interface between Tx line and load. The reflected wave then interacts with the original incident wave to produce a stationary wave. Thus at steady state, the ratio of voltages and currents along the distance ($l$) from load end of line does not remain constant and is given by \cite{Pozar}\cite{White}
	
	\begin{equation}
		Z_{in}(l)=Z_0\left[ \frac{Z_L+jZ_0tan(\beta l)}{Z_0+jZ_Ltan(\beta l)}\right] 
		\label{Impedance on line}
	\end{equation} 
	In a case where Tx line with characteristic impedance much higher than 50$\Omega$ is connected to a 50$\Omega$ load, a mismatch between them will cause the load impedance to be transformed into different values along the line at various positions ($l$). Under this condition, a special case arises where the length of the line equals to integral multiples of $\lambda$/2. Where $\lambda$ is wavelength. As per Eqn(\ref{Impedance on line}) at these wavelengths the load impedance does not transform but remains unchanged ($Z_{in}=Z_{L}$). Thus for a known value of the length of line ($l$), the wavelengths and the corresponding frequencies for which the impedance remains unchanged is given by
	\\
	\begin{equation}
		\lambda = \frac{2l}{n} \hspace{1cm}  f=\frac{v}{\lambda}= \frac{nv}{2l}; \hspace{0.5cm} n=1,2,3...
		\label{length dependence of wavelength}
	\end{equation}
	 $v$ is velocity of EM waves. This situation is generally avoided since almost all applications demand no transformation of load throughout the band of interest. But alternatively an effective use of this phenomenon can be envisaged to transmit power only at the wavelength where impedance remains untransformed. Here a 50$\Omega$ source will transmit maximum power only at frequencies corresponding to wavelengths given by Eqn(\ref{length dependence of wavelength}). While at other frequencies, because of mismatching, the power will simply reflect back to the source. If a purely resistive load is connected to the line, the impedance seen by the source contains non-zero reactance at all frequencies except for the frequencies at which condition of Eqn(\ref{length dependence of wavelength}) is satisfied. Thus if the load is purely resistive, resonances (zero reactance) can only be seen at frequencies of no transformation. 
	
	The inherent property of a mismatched Tx line to resonate only at a set of frequencies for a resistive load like 50$\Omega$ can help to realize a filter. The response of a mismatched Tx line is such that it is capable of transmitting signals only in a small bandwidth at the line's resonant frequencies. The transfer function of line resembles that of a narrow bandwidth bandpass filter which repeats itself after a specific interval in frequency domain. Fig(\ref{irregular shift in resonant freq}a) illustrates the response of a mismatched Tx line. The S-parameter response has been calculated in MATLAB using Eqn(\ref{Impedance on line}).
	
	The relation mentioned in Eqn(\ref{length dependence of wavelength}) can also be obtained by considering the occurrence of multiple reflections inside the mismatched Tx line. Similar to the case of Fabry-Perot(FB) Etalon, the incident power on the interface of mismatched Tx line is only partly transmitted into the line. The interfaces act like semi-transparent mirrors of FB etalon. The transmitted wave suffers yet another reflection at the other interface and some part of it travels backwards. This sets a series of reflections every time the wave is incident on any one of the interface.  Thus at steady state, an impedance profile as given by Eqn(\ref{Impedance on line}) is formed. But the frequencies at which a standing wave is created inside the mismatched line, the impedance does not transform giving maximum transmission.  
	
	\subsection{Tuning the Response of Mismatched Tx Line}
	It is clear from formula (\ref{length dependence of wavelength}) that the frequencies of resonance depend upon length of Tx line ($l$) and velocity ($v$). Thus in order to tune the frequencies of resonance both $v$ and $l$ can be varied. 
	In the present study we concentrate on dependence on velocity for a fixed length. It is well known from the theory, and stated in Eqn(\ref{Char impedance and prop const}c) that the phase velocity of waves depend upon the Capacitance per unit length (C) and Inductance per unit length (L). Thus by altering the dielectric material between the pair of conductors, the capacitance per unit length can be varied and hence velocity of waves can be adjusted in such a way that the resonance occurs at the desired value. In fact the change in the velocity manifests as change in the wavelengths of frequencies on the line. Hence if the physical length ($l$) of the line is kept fixed; the condition of Eqn(\ref{length dependence of wavelength}) gets satisfied at a different set of frequency values. \\As dielectric constant of frequently used materials is greater than that of air, introduction of any dielectric will only reduce the velocity of waves. Hence wavelengths of all frequencies are shortened and under the condition of Eqn(\ref{length dependence of wavelength}), resonant frequencies will be lesser than earlier. Thus the variation in first resonant frequency is a function of relative dielectric constant ($\epsilon_r$) and as suggested by Eqn(\ref{length dependence of wavelength}), it will have an inverse relation with  $\sqrt{\epsilon_r}$. 

\begin{figure}[h]
	\begin{subfigure}{0.3\textwidth}
		\includegraphics[width = 5cm]{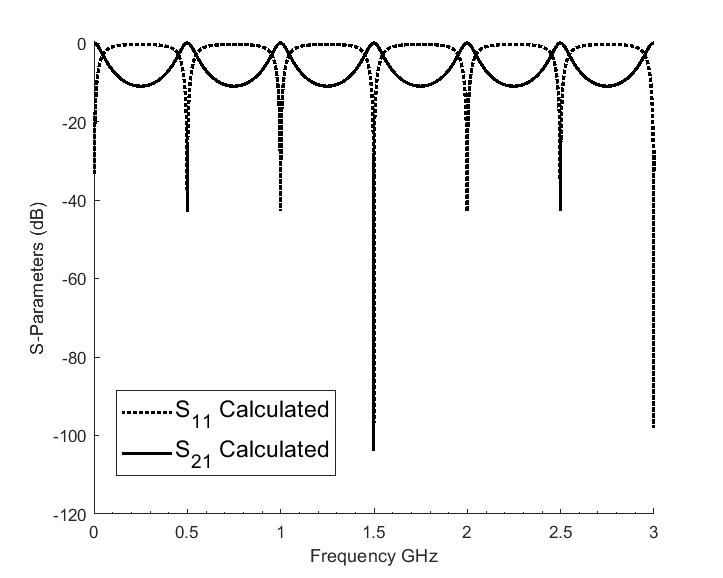}
		\caption{}
	\end{subfigure}
	\begin{subfigure}{0.3\textwidth}
		\includegraphics[width = 5cm]{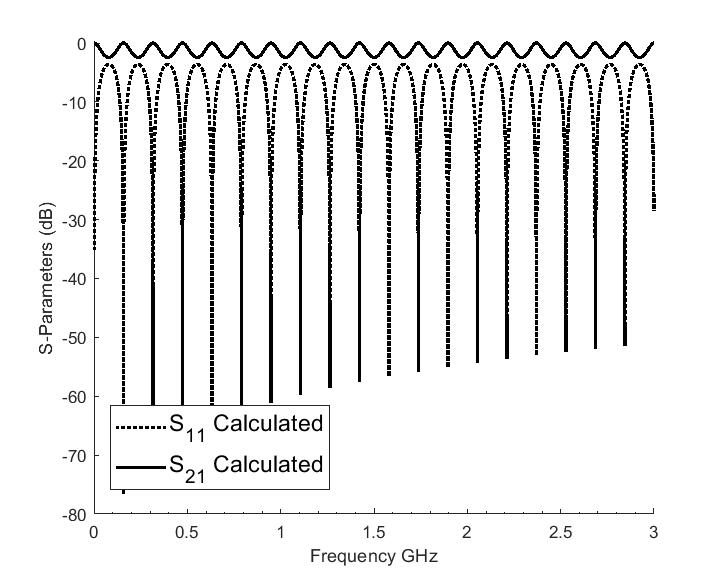}
		\caption{}
	\end{subfigure}
	\begin{subfigure}{0.3\textwidth}
		\includegraphics[width = 5cm]{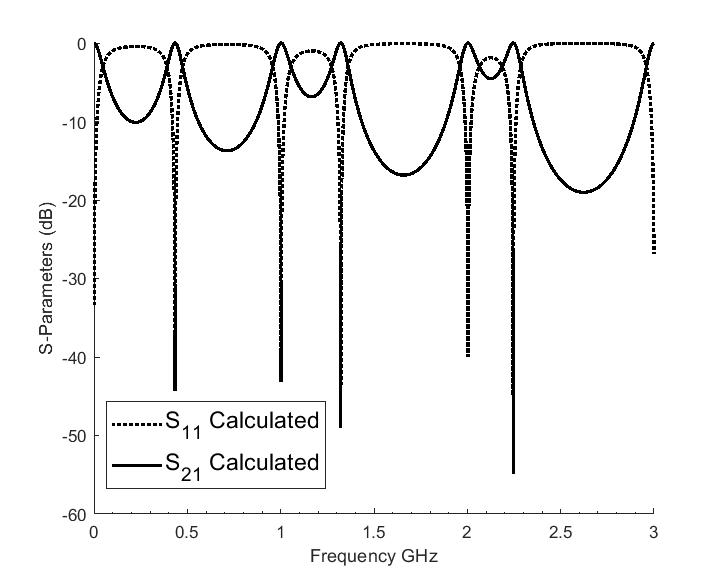}
		\caption{}
	\end{subfigure}
	\caption{A 30 cm long mismatched parallel wire Tx line ($Z_0=347\Omega$) has been considered. S parameters have been calculated using Eqn(\ref{Impedance on line}). (a) shows the resonant frequencies when $\epsilon_r$ is 1. (b) shows resonant frequencies when $\epsilon_r$ is changed to 10 for the entire line. (c) shows the resonant frequencies obtained when the $\epsilon_r$ is changed to  10 for a small length near the center of the line while rest of the line has $\epsilon_r$= 1.  Notice that due to reduction in mismatch the overall transmission in (b) has risen where as when the dielectric was changed locally in (c), the desired pattern is preserved although shift in the resonant frequencies is non-uniform.}
	\label{irregular shift in resonant freq}
\end{figure}

	 So just by filling the gap between the conductors of Tx lines with a dielectric of suitable value a new set of resonating frequencies can be obtained. The approach mentioned here is very similar to what is adopted by microfludic tunable electronic devices wherein the channels created in the devices are filled with fluid dielectrics. The key difference, however, is the use of mismatched Tx line to generate the bandpass filter like transfer characteristics. But the drawback of replacing the dielectric through out the length of line is that the characteristic impedance of the Tx line also drops closer to 50$\Omega$(as it too depends inversely on $\sqrt{\epsilon_r}$ ) and hence the mismatch between the line and a 50$\Omega$ system reduces which effectively raises transmission coefficient at all frequencies. Fig(\ref{irregular shift in resonant freq}b) illustrates the case. Here the Eqn(\ref{Impedance on line}) was solved to obtained S-parameters. One of the possible alternative is discussed in section \ref{achieving tunability}. 

	\subsection{Tuning with Localized Dielectric Variations}
	\label{achieving tunability}
\begin{figure}[h]
		\centering
		\includegraphics[scale=0.65]{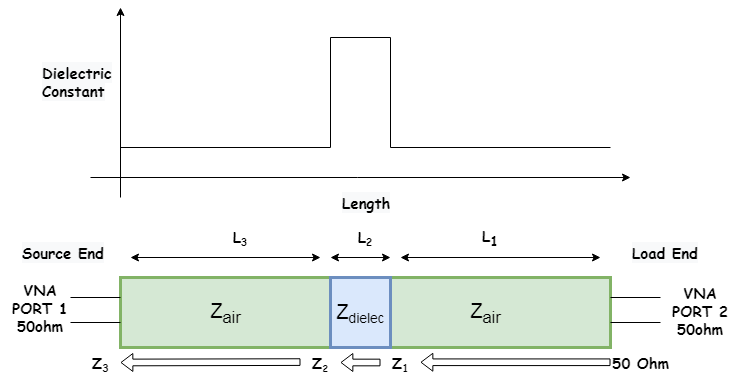}
		\caption{Dielectric profile and impedance transformation on the line}
		\label{dielectric profile}
	\end{figure}

	Apart from raising the transmission throughout the band as illustrated in Fig(\ref{irregular shift in resonant freq}b), it is often inconvenient to change the entire dielectric between the lines as it would demand a well developed mechanism to hold and control the dielectric flow \cite{IP}. The dielectrics may not always be available in a form which is easily employable. The amount of the dielectric can also be a concern. Finally the change in the resonant frequency will only occur in steps governed by the values of dielectric constant of easily employable materials. Hence to address these impediments, a proof of concept has been experimentally shown in the current study that it is not always necessary to change the dielectric in the entire line. Instead it can be changed locally on the line (as shown in Fig(\ref{dielectric profile})) to obtain similar effect. A discontinuity in the dielectric profile of the line  alters the velocity of EM waves. A window shaped profile of dielectric constant along the line is resulted due to local change in the dielectric medium. The region of line coinciding with the window in dielectric profile would have different value of characteristic impedance and propagation constant than the remaining line. This perturbs the impedance profile on the line. The entire scenario can be modeled as three Tx lines with different characteristics in cascade as shown in Fig(\ref{dielectric profile}).
	The total length of the line is divided into three sections. The impedance connected to load end of Tx line (near VNA port 2) is transformed into some value of impedance (say) Z1. Now, Z1 acts as a load to the second section (L2) of the Tx line further transforming it to Z2. And finally Z2 is transformed into Z3 at the source end (VNA Port 1) because of the third section (L3) of the Tx line.  Since the output impedance of port 1 is 50$\Omega$, the resonance happens for a frequency at which Z3 equals 50$\Omega$.
	
	Although the method gives a tighter control over tuning individual resonant frequency through out the band of observation, it also gives counter-intuitive changes in higher resonant frequencies.  Because of the limited length of the second section (L2) of Tx line, depending upon its position throughout the line, resonant frequencies of higher order change in an irregular manner. This is illustrated in Fig(\ref{irregular shift in resonant freq}c). 
	Many different possibilities of dielectric profiles can be studied in order to get better control over the phenomenon.	
	\section{Experimental}
	\label{Observations}
\begin{figure}[h]
		\begin{subfigure}{0.3\textwidth}
			\includegraphics[width = 5cm]{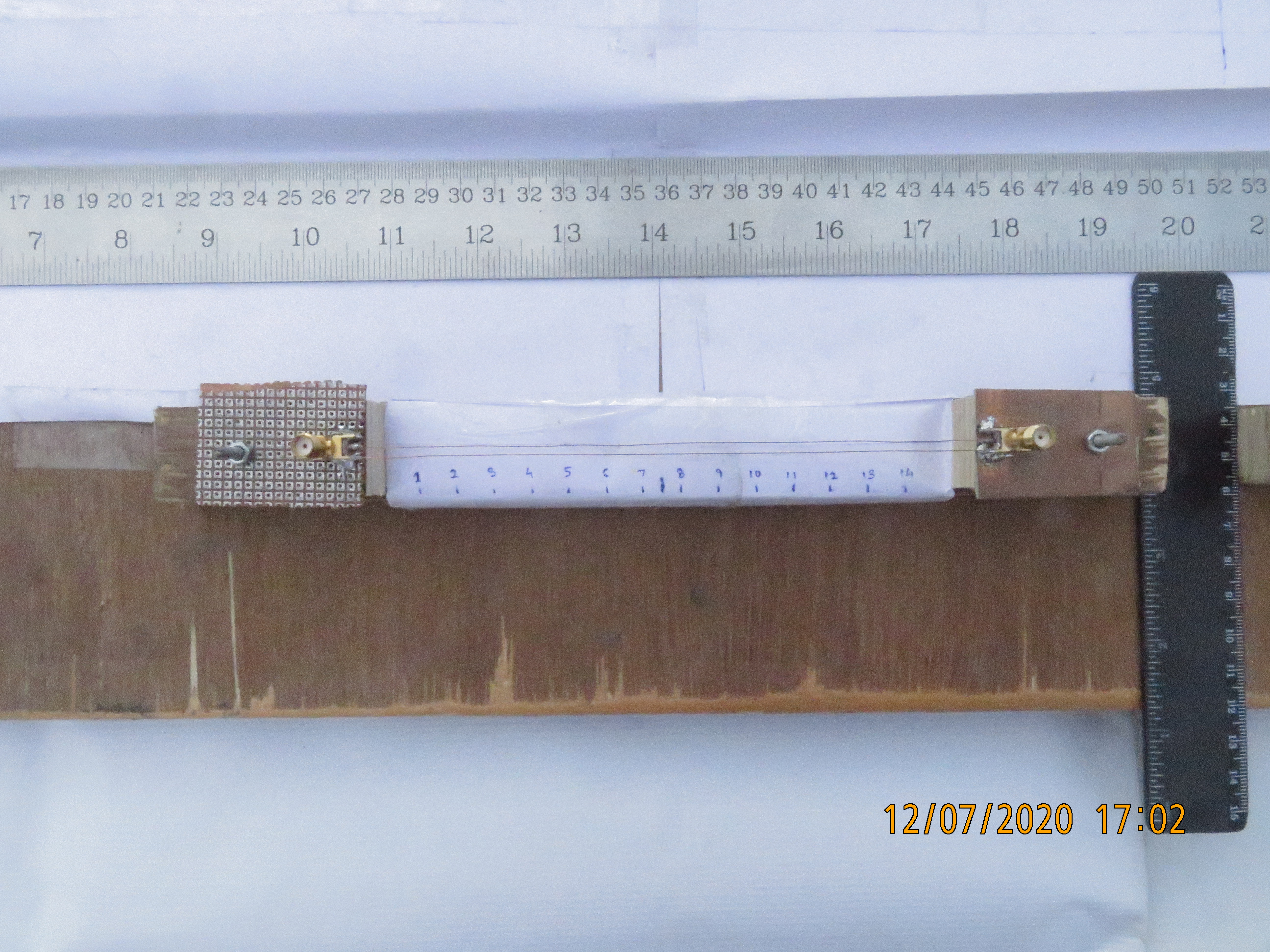}
			\caption{Length = 15cm}
		\end{subfigure}
		\begin{subfigure}{0.3\textwidth}
			\includegraphics[width = 5cm]{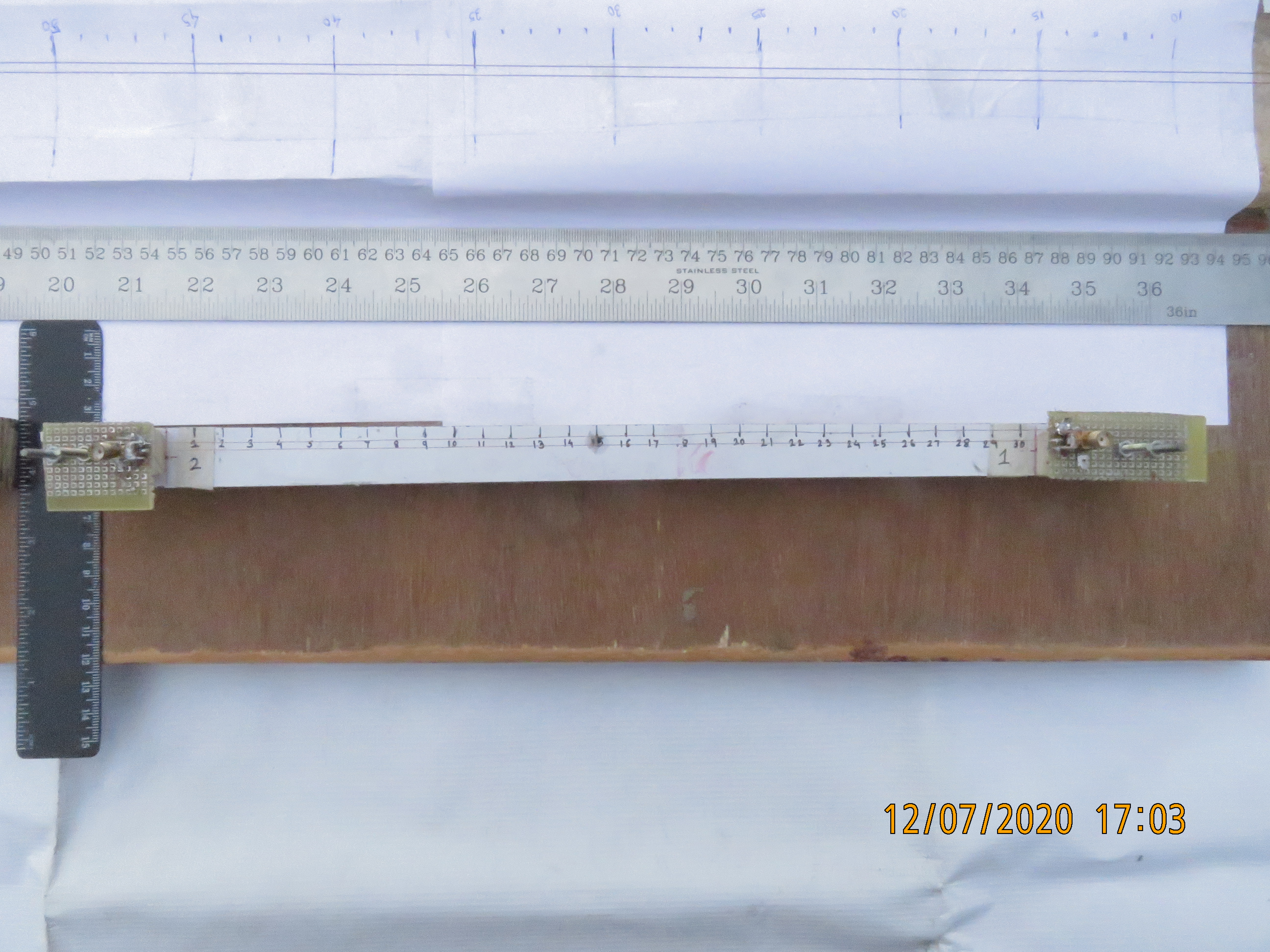}
			\caption{Length = 30 cm}
		\end{subfigure}
		\begin{subfigure}{0.3\textwidth}
			\includegraphics[width = 5cm]{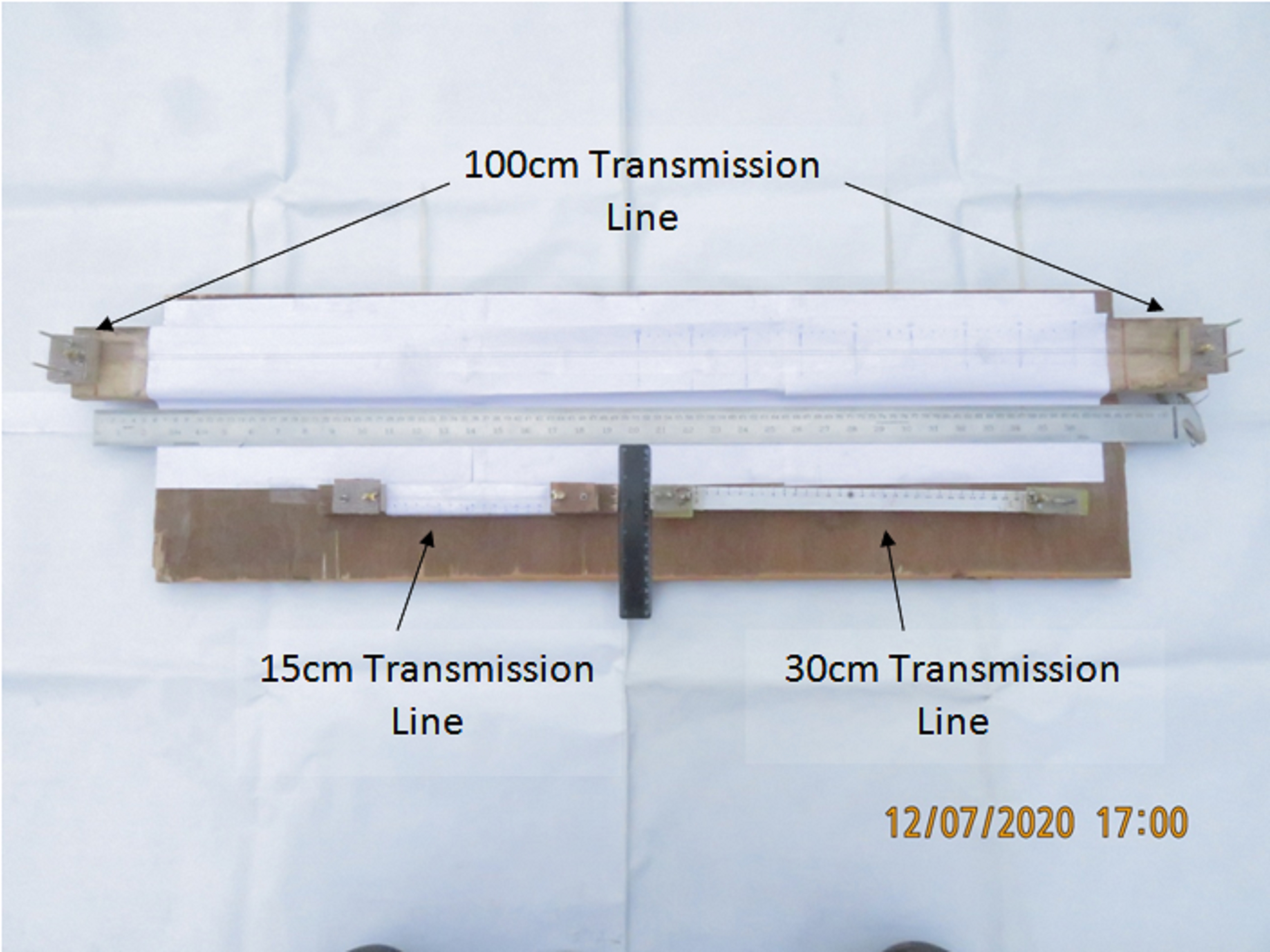}
			\caption{Length = 100 cm}
		\end{subfigure}
		\caption{Fabricated Tx Lines}
		\label{device image}
	\end{figure}

	It was experimentally verified that resonant frequency depends upon the choice of dielectric and of its position on the line.  Among various types of Tx lines, a parallel wire line is one of the most easy to fabricate and work with. Hence the study was performed on a parallel wire Tx line. The characteristic impedance of a parallel wire Tx line can be derived from Eqn(\ref{Char impedance and prop const}) for a practical lossless line and it is given by \cite{Pozar} Eqn(\ref{Characteristic impedance of parallel wire})
	\begin{equation}
		Z_{0}=\frac{276}{\sqrt{k}}log\frac{d}{r}
		\label{Characteristic impedance of parallel wire}
	\end{equation}
	
	  where $r$ is radius of wires and $d$ is separation between them. As shown in Fig(\ref{device image}), Tx lines of three different lengths consisting of two enameled copper wires of 110 micron radius with 2mm separation were fabricated. The wires were connected on to 50$\Omega$ SMA connectors on both the ends. For air dielectric, the characteristic impedance for the said configuration of the Tx line is 347$\Omega$. The calculated value is significantly higher than 50$\Omega$ and hence it can substantially manifest the phenomenon being tested. The entire setup of Fig(\ref{device image}) was mounted on a wooden base. As a thumb rule the conductors were mounted at a height which is at least ten times greater than the separation between the conductors to avoid disturbance by the wooden base to the EM fields produced by the lines.
	Three lines of lengths 15cm, 30cm, 100cm were fabricated to study the dependence of resonant frequency on length. It was made sure that the characteristic impedance of all the three lines was equal (separation $d$ was kept same). Three liquid dielectric materials viz. Ethanol ($\epsilon_r=25$) \cite{Nia}, Glycerol($\epsilon_r=50$) \cite{Schneider} and Water($\epsilon_r=78$) \cite{Kaatze} were chosen to test the phenomenon. 
	All the calculations and plots in current study were performed using MATLAB.\\ 

	\subsection{ Method of Dispensing Fluid Dielectrics Over Line}
	\label{placing Dielectric}
	Fluids have a tendency to flow away when a droplet is placed between the parallel conductors. This was a problem for the present study and hence it created a need of an agent which can hold all fluid dielectrics in place without significantly affecting the measurements being done.\par
\begin{figure}[H]
		\centering
		\begin{subfigure}{0.45\textwidth}
			\includegraphics[height=5cm, width = 7cm]{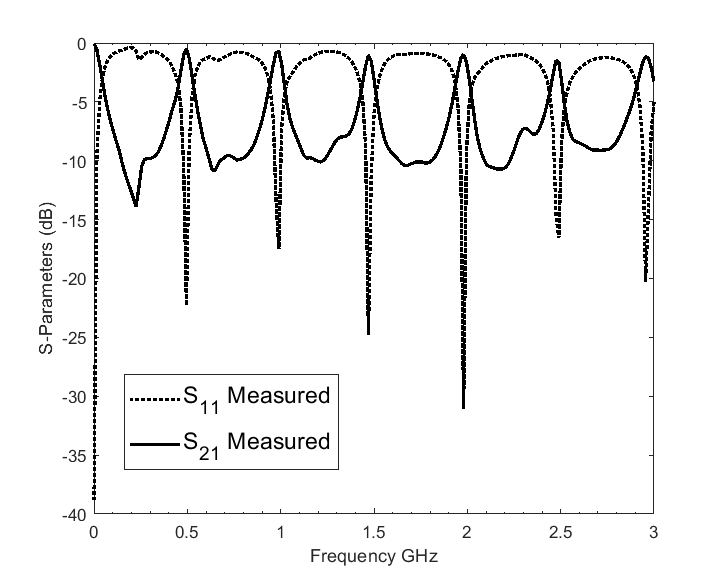}
			\caption{}
		\end{subfigure}	
		\begin{subfigure}{0.45\textwidth}
			\includegraphics[height=5cm, width = 7cm]{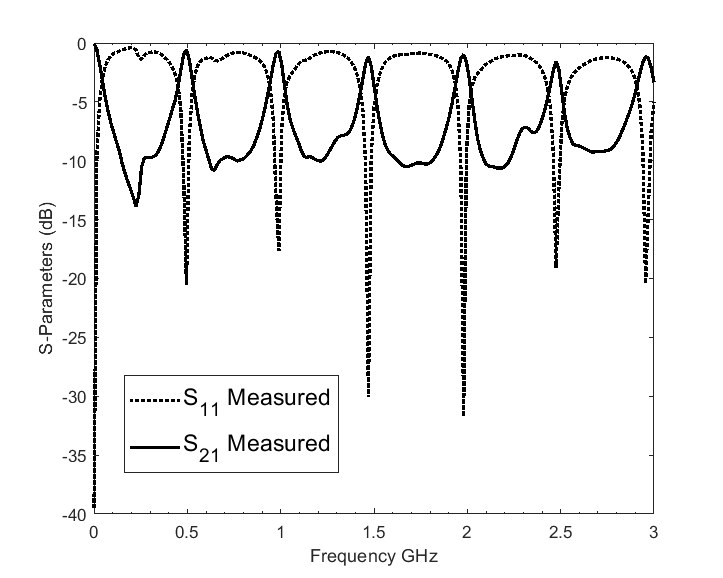}
			\caption{}
		\end{subfigure}	
		\caption{(a) shows the S-parameters without the piece of filter paper. (b) shows S-parameters with the piece of filter paper.}
		\label{effect of filter paper}
	\end{figure}

\begin{figure}[H]
		\centering
		\begin{subfigure}{0.45\textwidth}
			\includegraphics[height=5cm, width = 7cm]{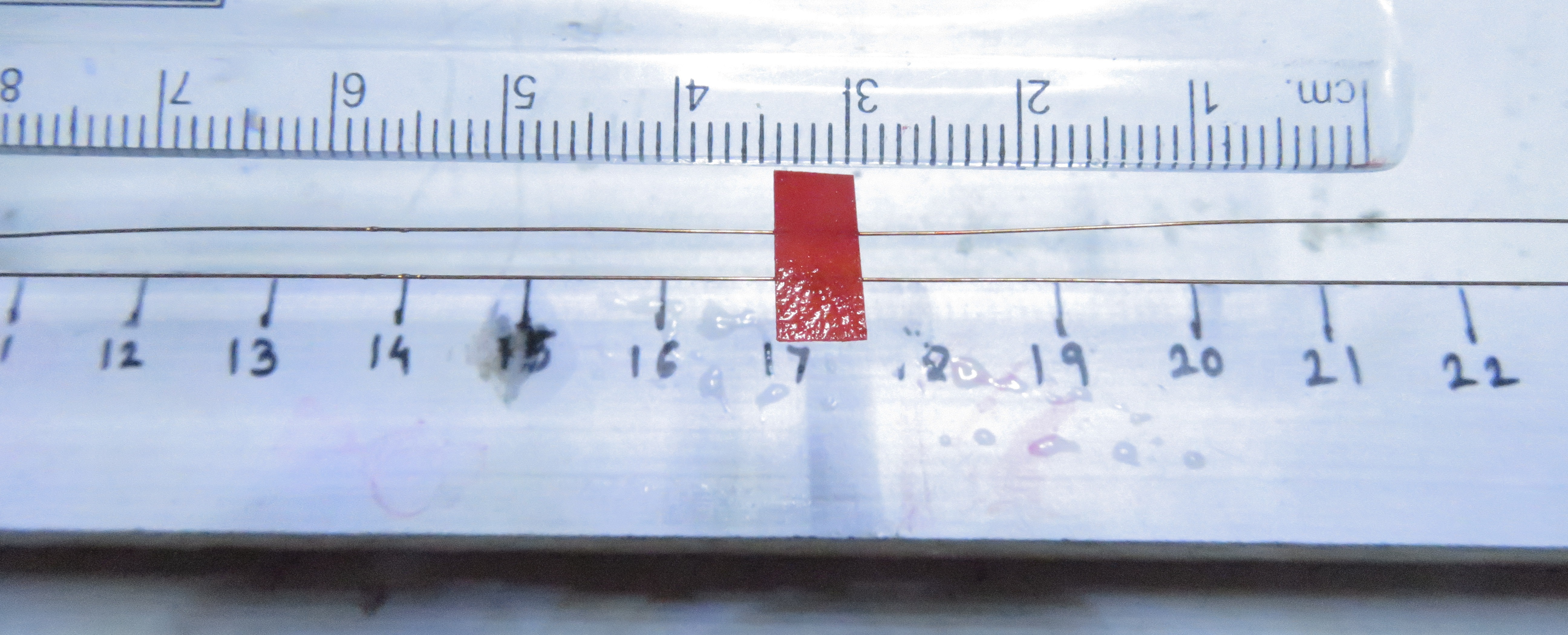}
			\caption{}
		\end{subfigure}
		\begin{subfigure}{0.45\textwidth}
			\centering
			\includegraphics[height=5cm, width = 7cm]{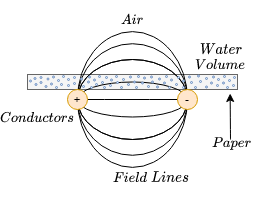}
			\caption{}
		\end{subfigure}
		\caption{ (a) shows a piece of filter paper saturated with the liquid dielectric (here dipped in red ink) which has been kept on the line. (b) show the field lines partially residing the water volume. The effective dielectric constant arising from the structure should be used to calculate S-parameters.}
		\label{paper on line}
	\end{figure}

 
	It was found that a grade 40 Whatman filter paper can be at disposal. A small piece of it (10mm $\times$ 5mm) can be balanced on the parallel conductors. In principle, this extra material kept on the conductors will have an effect on the properties of the line. Practically, however, the measuring instrument used is insensitive to detect these minuscule variations. On a coarser scale no effect due to the presence of paper was observed on resonant frequencies for all three lengths. Also the effect on resonant frequencies due to liquid dielectric is far more significant than the effect caused due to the piece of paper. Plots in Fig(\ref{effect of filter paper}) show the S parameters with and without the paper in place on a 30 cm long Tx line. All the resonant frequencies almost coincides with each others.
	In order to introduce the dielectric between the lines the filter paper was dipped in the liquid dielectric and then placed on the lines. In this case due to surface tension, the liquid quickly adheres to both copper wires. Thus the formed bridge of dielectric is sufficient to modify the electric field lines in the region.  Fig(\ref{paper on line}a) shows the piece of filter paper dipped in red color ink to illustrate the technique of placing it over the lines. It is important to note that the model proposed in section \ref{achieving tunability} tries to explain the behavior of the lines and the phenomenon being studied \emph{at the level of first order approximation.}
	The model \emph{ignores} the fact that dielectric is not present throughout the space in up-down and lateral directions; but is confined to the filter paper. Thus as opposed to ideal case in which field lines emanating from conductors will reside entirely in dielectric, the field lines in case of filter paper will partly reside in the dielectric (Fig\ref{paper on line}b). Since the volume occupied by the filter paper (and thus the dielectric material) is very less, the effective dielectric constant produced by the structure will be smaller than the actual value. Thus in order to validate the proposed model with the measurements, the effective dielectric constant arising from the structure should be chosen. 
	
	\subsection{Description of Experiments}
	 The phenomenon mentioned earlier has been tested through three experiments. All the measurements were performed on Agilent's E5062A vector network analyzer with the reference impedance of 50$\Omega$. Here the specific experimental conditions for all the three experiments have been given. The results obtained from the experiments have been discussed in detail in section \ref{results}
	 
	 \paragraph{Experiment A}
	 The aim of the experiment is to describe the characteristics of mismatched Tx line. In this experiment, the response of the Tx line was recorded with air as dielectric between the lines. 
	 \paragraph{Experiment B}
	 The experiment is aimed at investigating the possibility of tuning the center frequency of the mismatched Tx line. In this experiment three liquid dielectrics were introduced on the line by using the method mentioned earlier. For this experiment it is necessary that the piece of filter paper should be saturated with the dielectric before placing on the Tx line. Saturation of the paper ensures that approximately same volume of the liquid is introduced between the lines each time the experiment is repeated. Also it ensures that the presence of dielectric is dominant over filter paper. With the method used, there are other factors such as evaporation of dielectric liquids which can affect the measurements. It was thus ensured that the delay between placing the dielectric and recording the data was kept minimum. 
	\paragraph{Experiment C}
	The aim of the experiment is to observe the effect of volume of liquids on resonant frequency and to explore the utility of an assembly similar to the used experimental setup as a mechanism for sensing variations in small volume of liquid dielectrics. Since the effective dielectric constant of the structure in Fig(\ref{paper on line}b) depends upon the volume occupied by the water, it is suggestive that variation in the volume would change the effective dielectric constant of the structure. Hence even if the location of the filter paper is kept fixed, increasing the water content on the filter paper would decrease the resonant frequency. This was experimentally checked with use of Water on a 30cm long Tx line. A dry piece of filter paper was first kept on the Tx line at its center and then a fixed amount of water was dispensed on to the paper using a Hamilton's 5$\mu$L syringe. The shift in the resonant frequency was noted.  The water volume was increased in steps of 2$\mu$L. After placing the water on the filter paper, it was allowed to evaporate completely. As the water slowly evaporates the resonant frequency returns to its original value. The time required for resonant frequency to return to its original value was recorded. This is the total evaporation time under the condition of experimental setup. For each volume of water the process was repeated 3 times to get the average of all observations. The measurements were performed at a room temperature of 23\si{\degreeCelsius} and in an air conditioned closed environment  to ensure non variance of parameters like humidity, air flow and temperature as they affect the rate of evaporation of liquids.

	\section{Results and Discussion}
	\label{results}
	\subsection{Experiment A: Tx Lines Without Presence of the Dielectric Liquids}
\begin{figure}[h]
		\begin{subfigure}{0.3\textwidth}
			\includegraphics[width = 5cm]{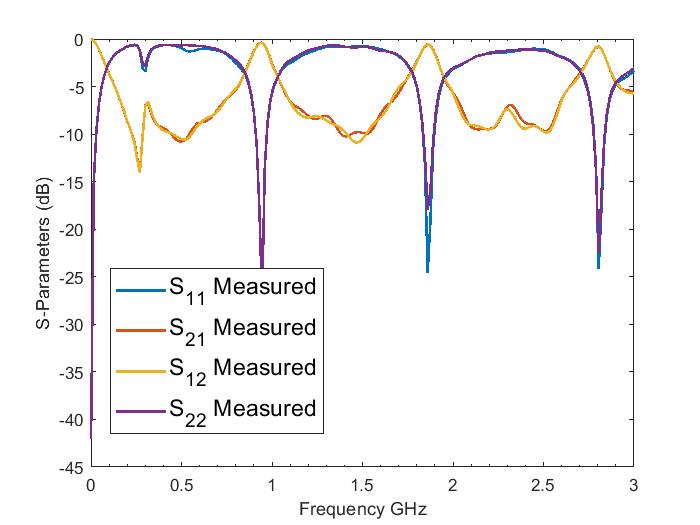}
			\caption{Length = 15cm}
		\end{subfigure}
		\begin{subfigure}{0.3\textwidth}
			\includegraphics[width = 5cm]{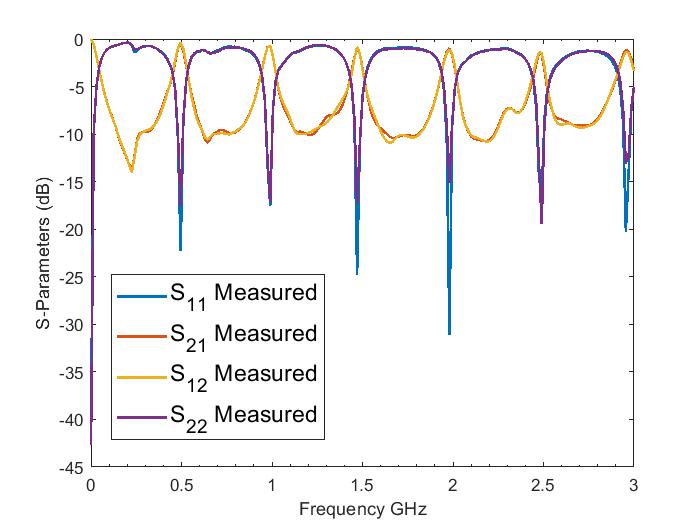}
			\caption{Length = 30 cm}
		\end{subfigure}
		\begin{subfigure}{0.3\textwidth}
			\includegraphics[width = 5cm]{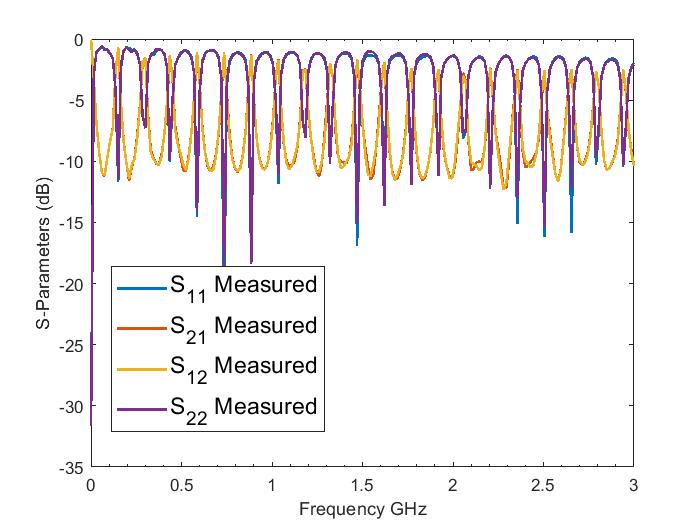}
			\caption{Length = 100cm}
	 	\end{subfigure}
 	\caption{Frequency response of the Tx line for three lengths}
		\label{S-para Only devices}
	\end{figure}

	As stated in section \ref{theory}, the resonant frequency of a mismatched line depends upon its length. Fig(\ref{S-para Only devices}) shows the S-parameters measured in a 50$\Omega$ system for three different lengths. Table(\ref{onlyDevice first three harmonics}) lists the expected values of the first three resonant frequencies (according to Eqn(\ref{length dependence of wavelength})) and the corresponding measured values.
	\begin{center}
	\begin{table}[h]
	\begin{tabular}{|c|c|c|c|c|c|c|}
		\hline
		\multirow{2}{4em} {Length} & \multicolumn{2}{|c|}{First Resonance} &  \multicolumn{2}{|c|}{Second Resonance} &  \multicolumn{2}{|c|}{Third Resonance}  \\
		\cline{2-7}
		  & Calculated & Measured & Calculated & Measured &Calculated & Measured \\
		 \hline
		 15cm & 1GHz & 945MHz & 2GHz & 1.86GHz & 3GHz & 2.80GHz \\
		 \hline
		30cm & 500MHz & 495MHz & 1GHz & 990MHz & 1.5GHz & 1.47 GHz \\
		\hline
		100cm & 150MHz & 150MHz	& 300MHz & 300MHz & 450MHz	& 435 MHz \\
		\hline
		
	\end{tabular}
		\caption{Measured and calculated values of first three resonances for all three Tx lines}
		\label{onlyDevice first three harmonics}
	\end{table}
	\end{center}

	It can be noticed from table(\ref{onlyDevice first three harmonics}) that there is a shift in the expected frequency of resonance. 
	 The shift is less at lower frequencies and it goes on increasing at higher frequencies. Also, the measured values of resonant frequencies are not harmonics in true sense, as they are not integral multiples of each other. In case of all fabricated lines, the S-parameters show unexpected behavior between two consecutive resonant frequencies. Although the exact reason behind this behavior is presently unclear, it is suspected to be an effect of method adopted to mount the SMA connectors. Since the lines are fabricated in-house, they may contain a few shortcomings such as imperfect soldering, irregularities in the distance between the wires, etc. Higher disagreement between the observed and calculated values at increasing frequencies could be an indication of irregularities of small dimensions, as they become comparable with the wavelength. Even though an exact match between the observed and calculated frequencies is missing, the Tx lines of all lengths follow the overall trend as explained in section \ref{theory}.

	It is known from Tx line theory that, the attenuation in line is an exponential function of real part of propagation constant $\alpha$. Dependence of $\alpha$ on R and G (from Eqn\ref{Char impedance and prop const}) represents ohmic losses and dielectric losses in the system. Taking only these into consideration, under mismatched conditions the ratio of power received by load $P_{load}$ to the input power given to the line $P_{in}$ is given by Eqn(\ref{PloadtoPin})(Refer to Appendix). The equation is valid only in case when the generator's impedance matches with the characteristic impedance of the line. In present study a 50$\varOmega$ generator is used to excite the Tx line of 347$\varOmega$ impedance. Hence the equation should be multiplied by $(1-{|\tau_{in}|}^2)$ resulting in Eqn(\ref{PloadtoPsource}) where $\tau_{in}$ is the reflection coefficient at the generator and Tx line interface.  This equation does not take into account the radiation losses due to the length of the Tx line. It was demonstrated in \cite{Reuven} that the radiation loss in case of a mismatched twin cable depends on forward traveling wave as well as on reverse traveling wave. Eqn(21) from \cite{Reuven} gives an expression for power loss relative to incident power. Using the expression for radiation loss, table(\ref{Losses in line}) gives the total loss as the addition of ohmic losses and radiation losses for first three resonances and compares it with the measured value.  
	
	\begin{sidewaystable}
	\begin{center}
	\begin{tabular}{|p{0.05\textwidth}|p{0.05\textwidth}|p{0.07\textwidth}|p{0.05\textwidth}|p{0.07\textwidth}|p{0.05\textwidth}|p{0.07\textwidth}|p{0.05\textwidth}|p{0.07\textwidth}|p{0.05\textwidth}|p{0.07\textwidth}|p{0.05\textwidth}|p{0.07\textwidth}|}
		\hline		
		\multirow{3}{4em}{Length}  & \multicolumn{4}{|c|}{First Resonance} & \multicolumn{4}{|c|}{Second Resonance} & \multicolumn{4}{|c|}{Third Resonance} \\
		\cline{2-13}
		& \multicolumn{3}{|c|}{Calculated @ 1GHz} & \multirow{2}{4em}{Measured @ 945 MHz} & \multicolumn{3}{|c|}{Calculated @ 2GHz} & \multirow{2}{4em}{Measured @ 1.86 GHz } & \multicolumn{3}{|c|}{Calculated @ 3GHz} & \multirow{2}{4em}{Measured @ 2.80 GHz}\\
		\cline{2-4} \cline{6-8} \cline{10-12}
		& Ohmic Loss & Radiation Loss & Total Loss & & Ohmic Loss & Radiation Loss & Total Loss & & Ohmic Loss & Radiation Loss & Total Loss &  \\
		\hline
		15 cm & 0.1555 dB & 0.0035 dB & 0.1576 dB & 0.40 dB & 0.2199 dB & 0.011 dB & 0.2284 dB & 0.55 dB & 0.2697 dB & 0.024 dB & 0.2890 dB & 0.75 dB \\
		\hline
		\multirow{2}{4em}{} & \multicolumn{3}{|c|}{Calculated @ 500MHz} & \multirow{2}{4em}{Measured @ 495 MHz} & \multicolumn{3}{|c|}{Calculated @ 1GHz} & \multirow{2}{4em}{Measured @ 990MHz} & \multicolumn{3}{|c|}{Calculated @ 1.5GHz} & \multirow{2}{4em}{Measured @ 1.47GHz}\\
		\cline{2-4} \cline{6-8} \cline{10-12}
		& Ohmic Loss & Radiation Loss & Total Loss & & Ohmic Loss & Radiation Loss & Total Loss & & Ohmic Loss & Radiation Loss & Total Loss &  \\
		\hline
		30 cm & 0.2191 dB & 0.0032 dB & 0.2196 dB & 0.40 dB & 0.3090 dB & 0.007 dB & 0.3112 & 0.69 dB & 0.3780 dB & 0.0137 dB & 0.3829 dB & 0.83 dB \\
		\hline
		\multirow{2}{4em}{} & \multicolumn{3}{|c|}{Calculated @ 150MHz} & \multirow{2}{4em}{Measured @ 150 MHz} & \multicolumn{3}{|c|}{Calculated @ 300 MHz} & \multirow{2}{4em}{Measured @ 300 MHz} & \multicolumn{3}{|c|}{Calculated @ 450 MHz} & \multirow{2}{4em}{Measured @ 435 MHz}\\
		\cline{2-4} \cline{6-8} \cline{10-12}
		& Ohmic Loss & Radiation Loss & Total Loss & & Ohmic Loss & Radiation Loss & Total Loss & & Ohmic Loss & Radiation Loss & Total Loss &  \\
		\hline
		100 cm & 0.3961 dB & 0.0081 dB & 0.3962 dB & 0.68 dB & 0.5560 dB & 0.016 dB & 0.5562 dB & 1.54 dB & 0.6774 dB & 0.024 dB & 0.6779 dB & 1.31 dB \\
		\hline
		
	\end{tabular}
		\caption{Loss in the Tx lines due to radiation as well as wire resistance}
		\label{Losses in line}
	\end{center}

\end{sidewaystable}

	It is clear from table(\ref{Losses in line}) that the radiation losses are very small in comparison with the other losses and hence may even be neglected in the further study. Notice that the frequencies for which the losses are calculated are slightly different than the frequencies of observation. This shift in the frequency is already discussed earlier. Calculated losses do not take into consideration the losses due to connector joints. Thus the difference  between measured and calculated losses can be ascribed to connector losses. Also it can be expected from the expression of ohmic losses that the longer lengths and higher frequencies should show more attenuation. This can be clearly seen in  Fig(\ref{S-para Only devices}c) as reduction in $S_{21}$ at higher frequencies in 100cm long Tx line as compared to other lines.

	 \subsection{Experiment B: Tx Lines with the Presence of Dielectric Materials}
\begin{figure}[hp!]
		
		\begin{subfigure}{0.3\textwidth}
			\includegraphics[width=5cm]{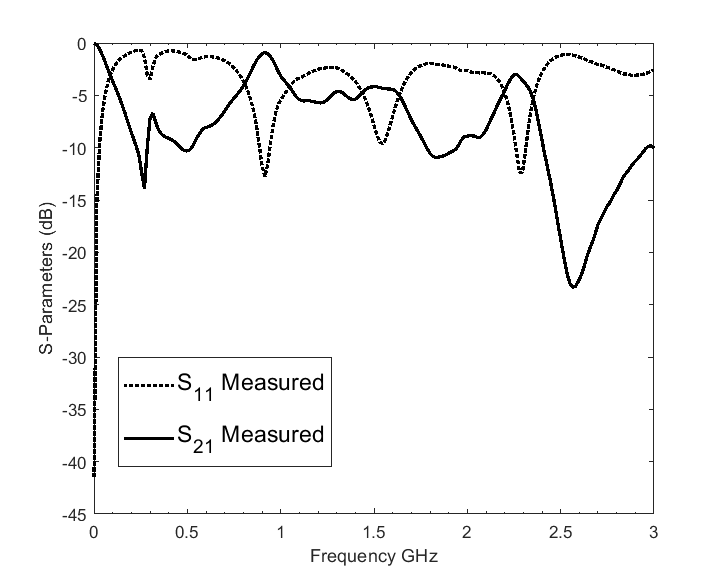}
			\caption{$x=1cm$}
		\end{subfigure}	
		\begin{subfigure}{0.3\textwidth}
			\includegraphics[width=5cm]{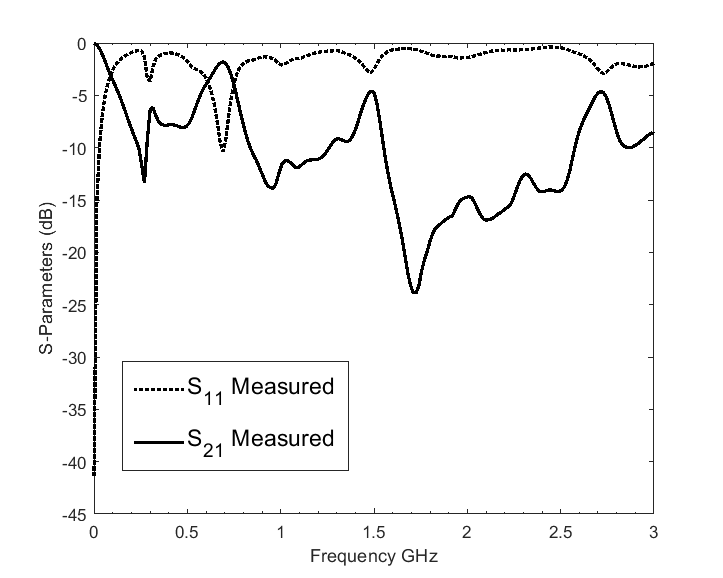}
			\caption{$x=4cm$}
		\end{subfigure}
		\begin{subfigure}{0.3\textwidth}
			\includegraphics[width=5cm]{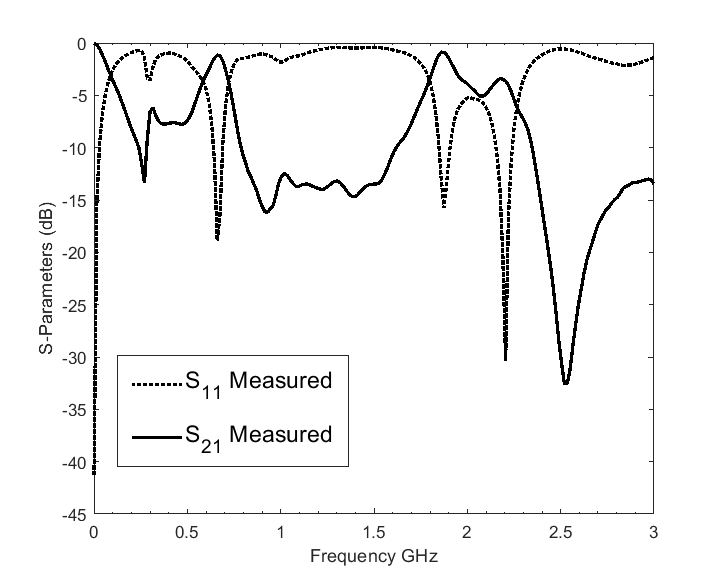}
			\caption{$x=7.5cm$}
		\end{subfigure}
		
		\begin{subfigure}{0.3\textwidth}
			\includegraphics[width=5cm]{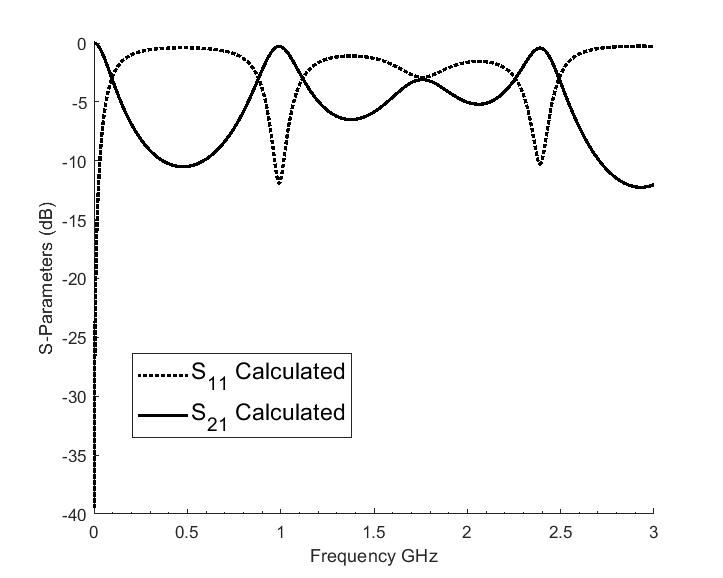}
			\caption{$x=1cm$}
		\end{subfigure}	
		\begin{subfigure}{0.3\textwidth}
			\includegraphics[width=5cm]{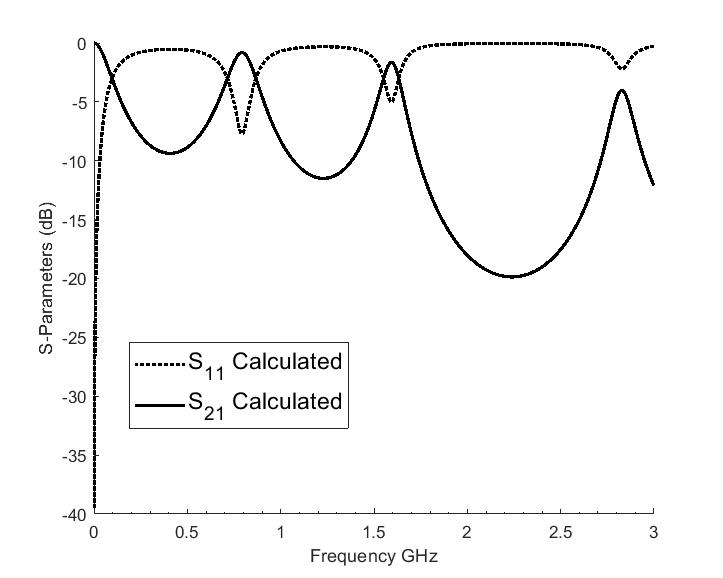}
			\caption{$x=4cm$}
		\end{subfigure}
		\begin{subfigure}{0.3\textwidth}
			\includegraphics[width=5cm]{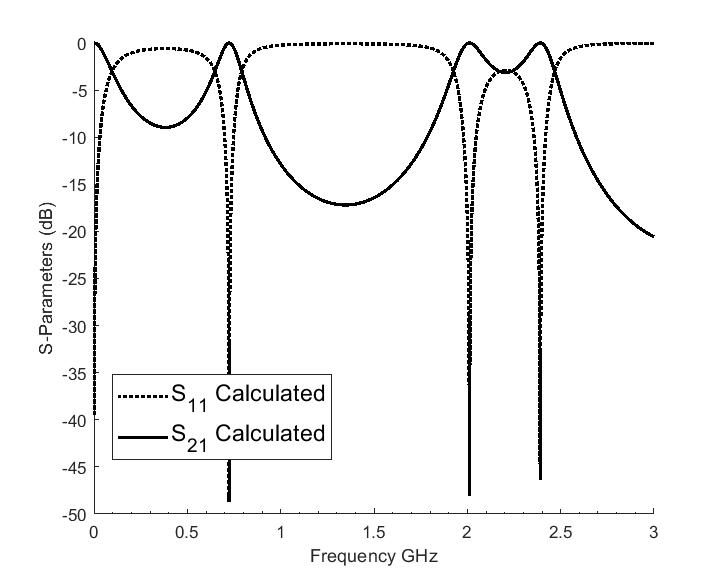}
			\caption{$x=7.5cm$}
		\end{subfigure}
		\caption{Water placed on a 15 cm Tx line at three different locations. The distance from the left end of the Tx line is denoted by $x$.  (a-c) show the measured S-parameters.  (d-f) show calculated S-parameters.}
		\label{Water on 15cm}	
	\end{figure}

\begin{figure}[hp!]
		
		\begin{subfigure}{0.3\textwidth}
			\includegraphics[width=5cm]{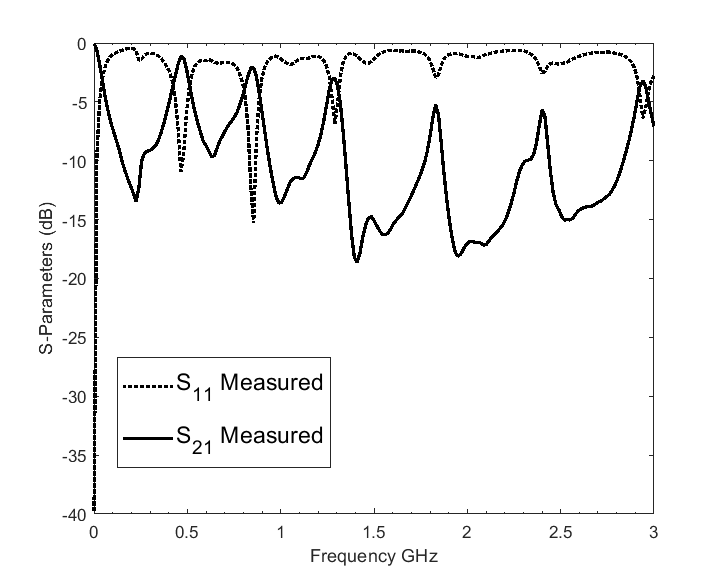}
			\caption{$x=5cm$}
		\end{subfigure}	
		\begin{subfigure}{0.3\textwidth}
			\includegraphics[width=5cm]{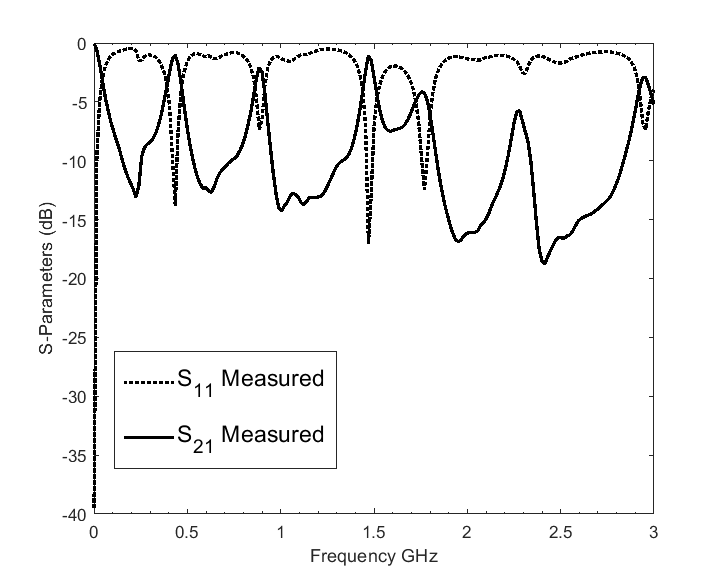}
			\caption{$x=10cm$}
		\end{subfigure}
		\begin{subfigure}{0.3\textwidth}
			\includegraphics[width=5cm]{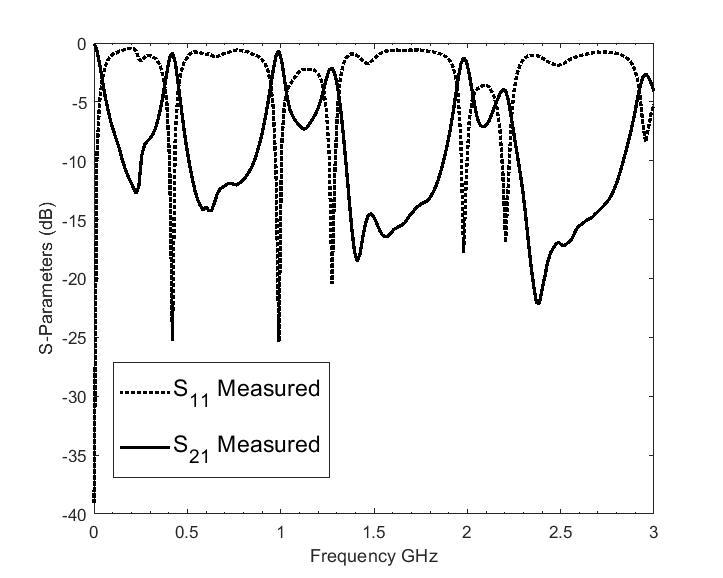}
			\caption{$x=15cm$}
		\end{subfigure}
		
		\begin{subfigure}{0.3\textwidth}
			\includegraphics[width=5cm]{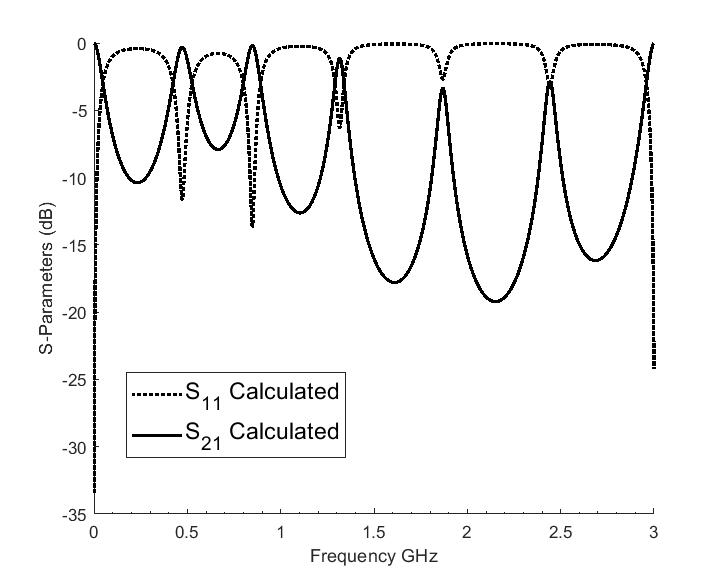}
			\caption{$x=5cm$}
		\end{subfigure}	
		\begin{subfigure}{0.3\textwidth}
			\includegraphics[width=5cm]{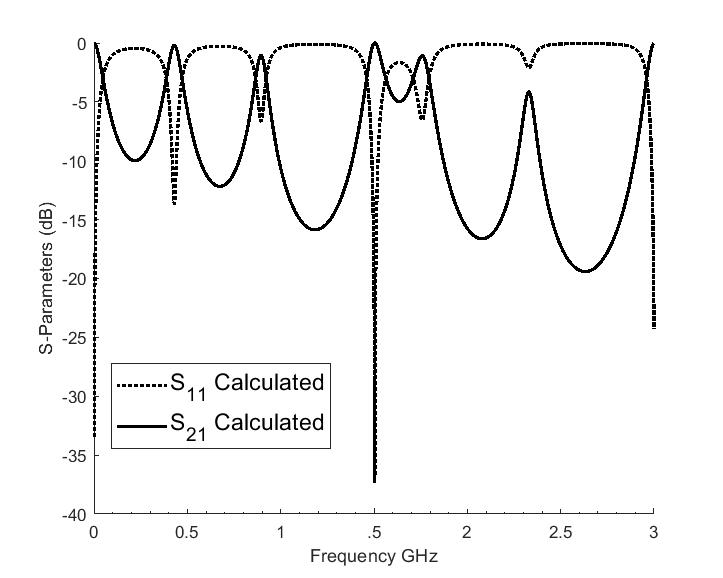}
			\caption{$x=10cm$}
		\end{subfigure}
		\begin{subfigure}{0.3\textwidth}
			\includegraphics[width=5cm]{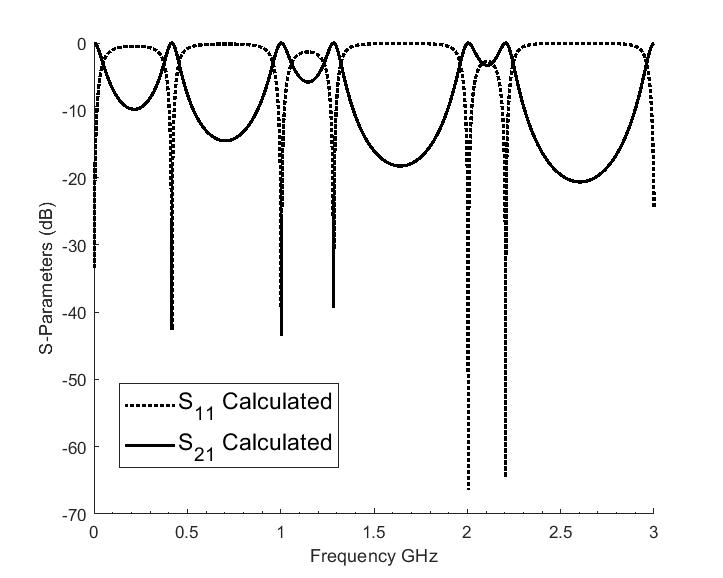}
			\caption{$x=15cm$}
		\end{subfigure}
		
		\caption{Water placed on a 30 cm Tx line at three different locations. The distance from the left end of the Tx line is denoted by $x$. (a-c) show the measured S-parameters.  (d-f) show calculated S-parameters.}
		\label{Water on 30cm}	
	\end{figure}

\begin{figure}[hp!]
		
		\begin{subfigure}{0.3\textwidth}
			\includegraphics[width=5cm]{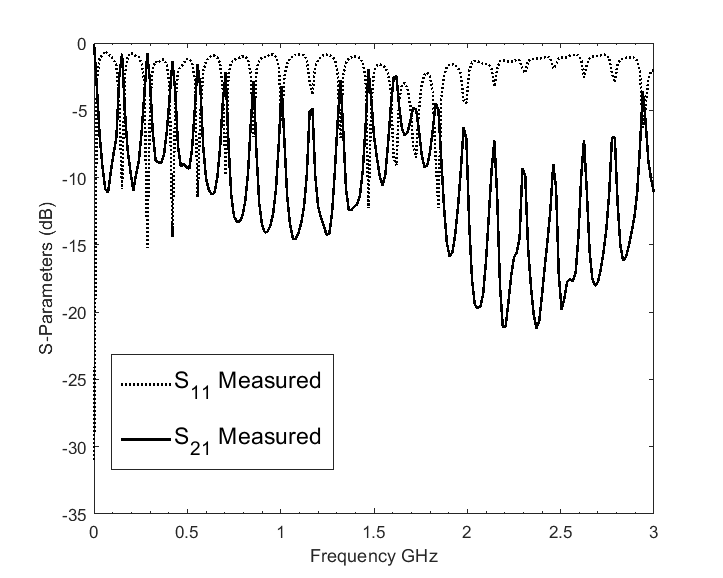}
			\caption{$x=10cm$}
		\end{subfigure}	
		\begin{subfigure}{0.3\textwidth}
			\includegraphics[width=5cm]{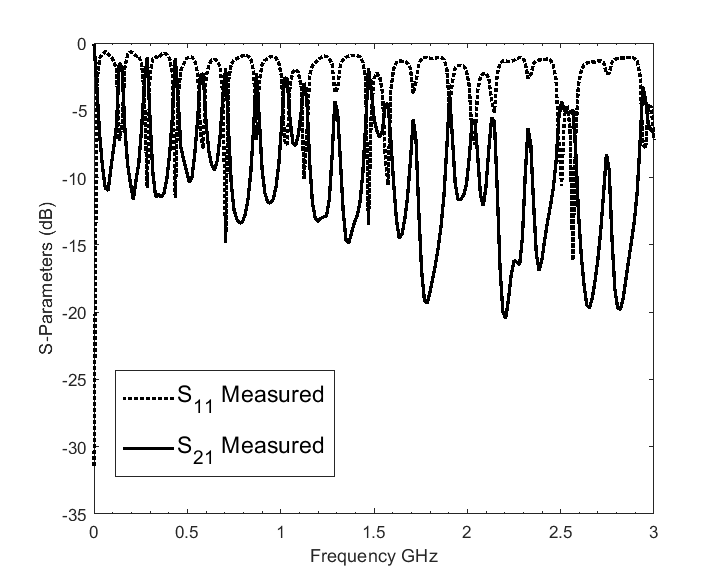}
			\caption{$x=30cm$}
		\end{subfigure}
		\begin{subfigure}{0.3\textwidth}
			\includegraphics[width=5cm]{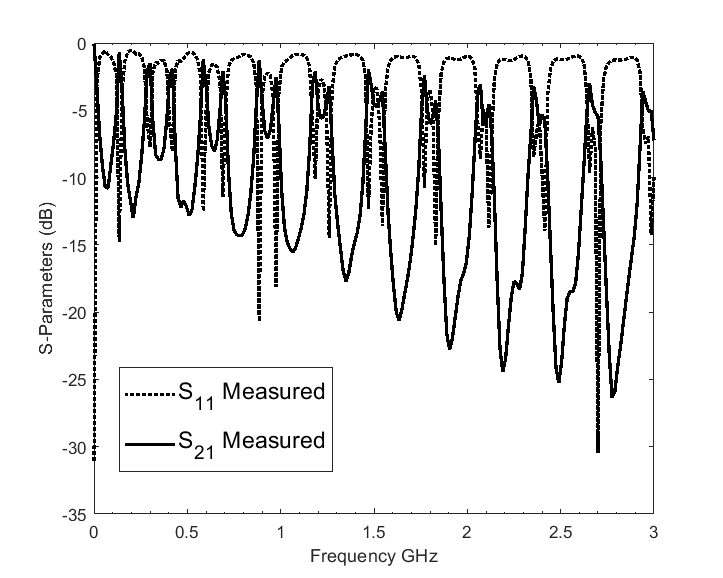}
			\caption{$x=50cm$}
		\end{subfigure}
		
		\begin{subfigure}{0.3\textwidth}
			\includegraphics[width=5cm]{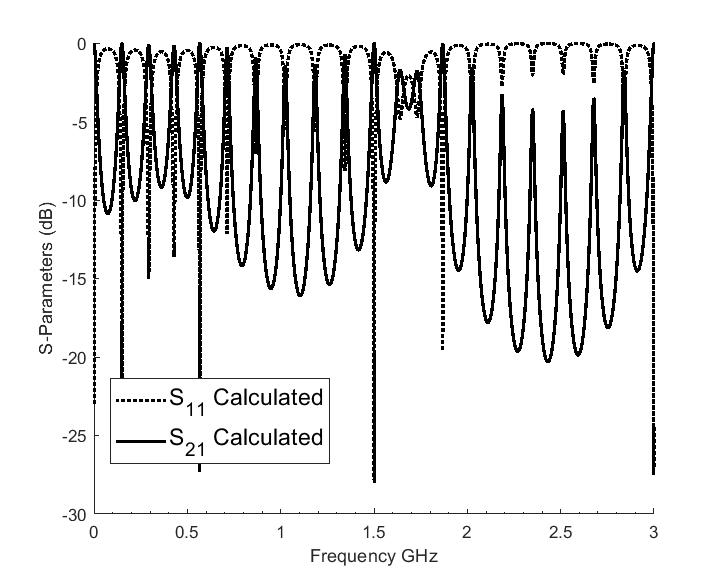}
			\caption{$x=10cm$}
		\end{subfigure}	
		\begin{subfigure}{0.3\textwidth}
			\includegraphics[width=5cm]{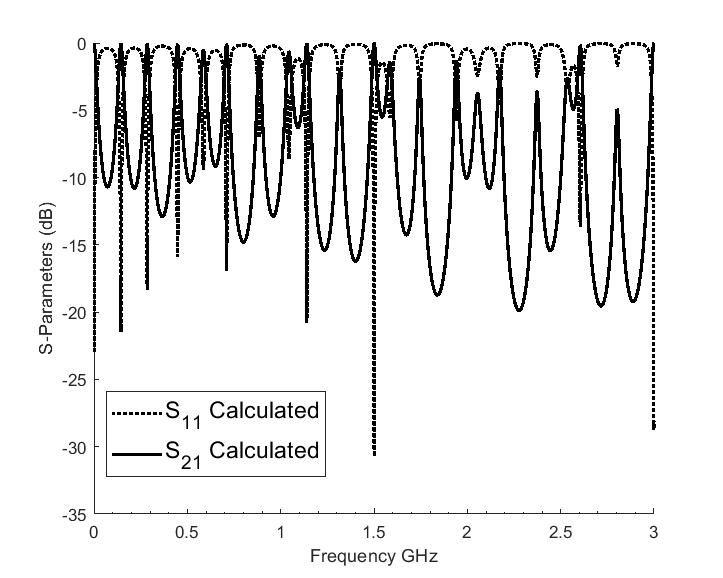}
			\caption{$x=30cm$}
		\end{subfigure}
		\begin{subfigure}{0.3\textwidth}
			\includegraphics[width=5cm]{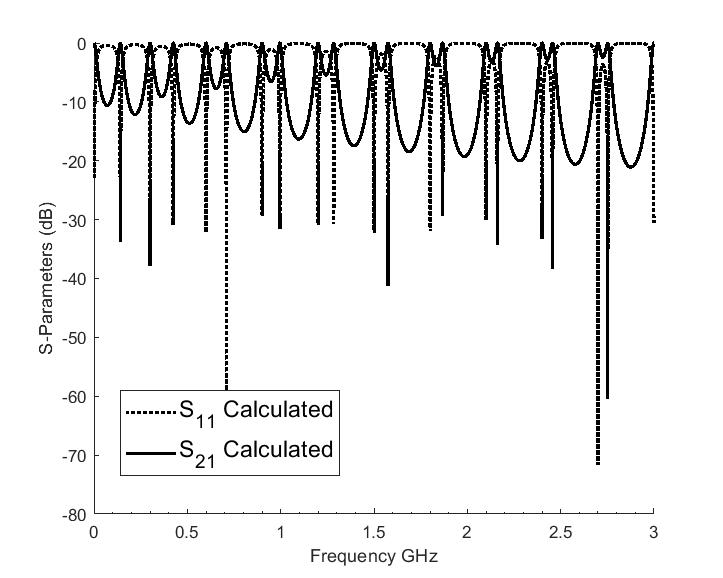}
			\caption{$x=50cm$}
		\end{subfigure}
		\caption{Water placed on a 100 cm Tx line at three different locations. The distance from the left end of the Tx line is denoted by $x$. (a-c) show the measured S-parameters. (d-f) show calculated S-parameters.}
		\label{Water on 1m}	
	\end{figure}

	  The materials with high dielectric constants are expected to show maximum change in resonant frequency when a local variation in dielectric profile is made. Tap water was chosen as one of the dielectric for present study as it has relatively high dielectric constant among readily available fluids. Due to Tx line's passivity and symmetry, it can be noticed from Fig(\ref{S-para Only devices}) that all measurements are reciprocal in nature  $(S_{11}=S_{22} \hspace{0.2cm}\& \hspace{0.2cm} S_{21}=S_{12})$. This means that \emph{only half the length} of Tx line is available for making changes in the dielectric profile. Hence the study was conducted only on the first half lengths of Tx lines. The repeatability in results was confirmed by taking 3 trials on each Tx line. Figures from (\ref{Water on 15cm}-\ref{Water on 1m}) show the obtained results. 
	 
	  As mentioned in section \ref{placing Dielectric}, the limited dimension of the filter paper reduces the effective dielectric constant of the region. The Eqn(\ref{Impedance on line}) can be used recursively to compute the S-parameters of the line by taking into account the reduced value of effective dielectric constant for section (L2). In figures from (\ref{Water on 15cm}-\ref{Water on 1m}), plots (a-c) show the measured results where as plots (d-f) show the computed values. It was found that the value of effective dielectric required to give a good agreement between measured and computed results is 13. Since the water is present near the lines occupying a very small volume, the effective dielectric has a large drop from 78 to 13.
	  Since calculation of S-parameters was done assuming a lossless system, the plots (d-f) do not take into account losses due to dielectric. The insertion loss in this case has an added component showing the dielectric losses. The loss tangents of dielectrics increases the conductance per unit length (G) in Eqn(\ref{Char impedance and prop const}) which further increases the value of $\alpha$ for section L2 in Fig(\ref{dielectric profile}) of the Tx line. The loss will be manifested as overall reduction in values of $S_{21}$ at various frequencies. The difference between the calculated plots and the measured plots of Fig(\ref{Water on 15cm}-\ref{Water on 1m}) can be due to the unmodeled losses in the dielectric.
	 
	 A peculiar result was observed in most of the plots at frequencies above 2GHz. Even though $S_{11}$ at some frequencies is significantly less ensuring no reflections, the $S_{21}$ plots at those frequencies show a strong dip, implying heavy signal losses. As the radiation losses in the system are insignificant (As shown in table(\ref{Losses in line})) the signals are getting absorbed in medium. This increase in losses at higher frequencies is in accordance with rising dielectric loss of water as previously reported in \cite{Lukenheimer}.

	 A unique characteristic of the mismatched lines can be noted from Fig(\ref{Water on 15cm}-\ref{Water on 1m}) that the response of lines towards dielectric change is more sensitive  near their center. Variation in the resonance frequency becomes insignificant as one approaches towards the end of Tx line. However in case of 1 meter long line, the shift in resonance frequency is still not noticeable at certain frequencies even though the dielectric was placed at the center of the line (Fig(\ref{Water on 1m}c and f)). This insignificant shift in the frequency of resonance at lower frequencies ($\leq$ 300 MHz) can be speculated due to the size of dielectric region being very small in comparison with the wavelengths at lower frequencies. 
	 
\begin{figure}[h!]
	 	\begin{subfigure}{0.3\textwidth}
	 		\includegraphics[width=5cm]{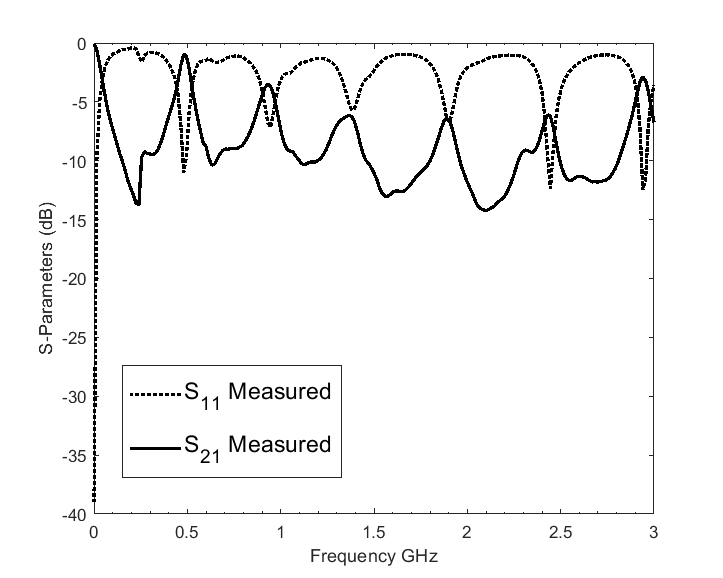}
	 		\caption{Ethanol @ $x=5cm$}
	 	\end{subfigure}
 		\begin{subfigure}{0.3\textwidth}
 			\includegraphics[width=5cm]{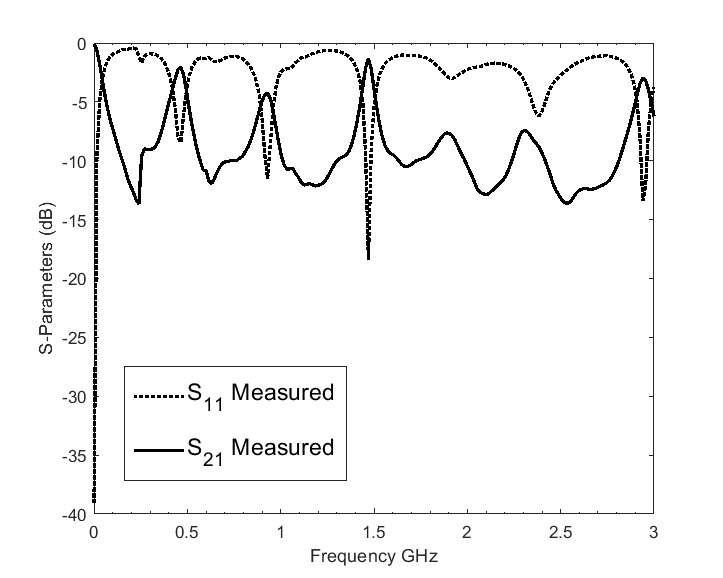}
 			\caption{Ethanol @ $x=10cm$}
 		\end{subfigure}
 		\begin{subfigure}{0.3\textwidth}
 			\includegraphics[width=5cm]{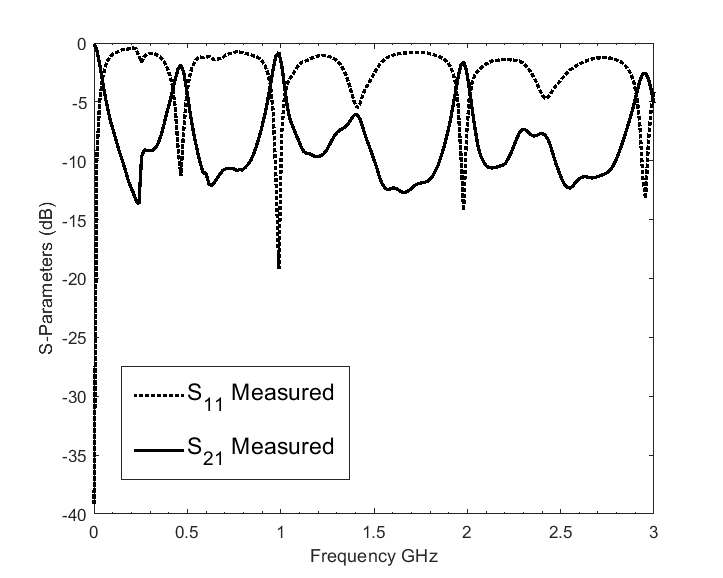}
 			\caption{Ethanol @ $x=15cm$}
 		\end{subfigure}
 		 \begin{subfigure}{0.3\textwidth}
 			\includegraphics[width=5cm]{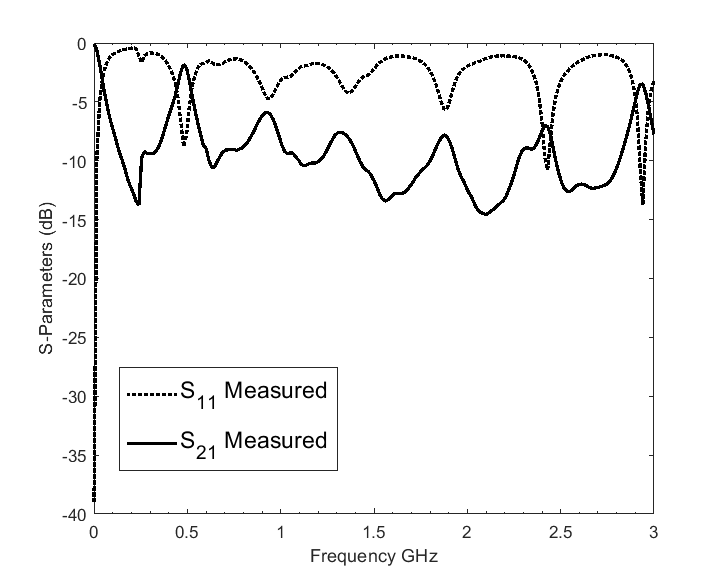}
 			\caption{Glycerol @ $x=5cm$}
 		\end{subfigure}
 		\begin{subfigure}{0.3\textwidth}
 			\includegraphics[width=5cm]{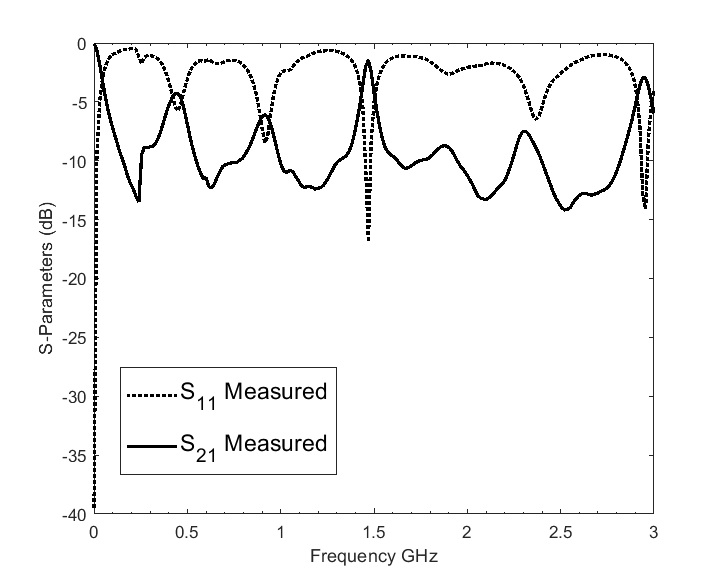}
 			\caption{Glycerol @ $x=10cm$}
 		\end{subfigure}
 		\begin{subfigure}{0.3\textwidth}
 			\includegraphics[width=5cm]{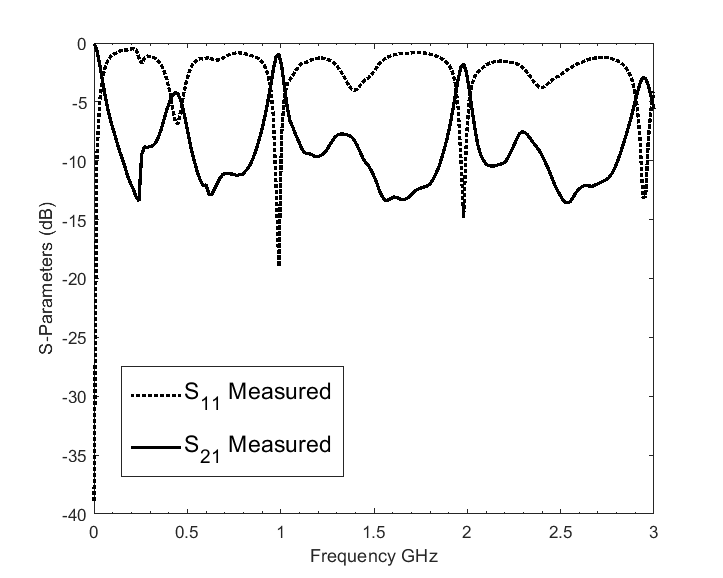}
 			\caption{Glycerol @ $x=15cm$}
 		\end{subfigure}
	 	\begin{subfigure}{0.3\textwidth}
 			\includegraphics[width=5cm]{Figs/Water/30cm/30CMPaperDr_5_Oneport.png}
 			\caption{Water @ $x=5cm$}
 		\end{subfigure}	
 		\begin{subfigure}{0.3\textwidth}
 			\includegraphics[width=5cm]{Figs/Water/30cm/30CMPaperDr_10_Oneport.png}
 			\caption{Water @ $x=10cm$}
 		\end{subfigure}
 		\begin{subfigure}{0.3\textwidth}
 			\includegraphics[width=5cm]{Figs/Water/30cm/30CMPaperDr_15_Oneport.png}
 			\caption{Water @ $x=15cm$}
 		\end{subfigure}
			\caption{Effect of dielectrics on the resonant frequency of mismatched Tx line. The dielectrics are placed at three different locations on a 30 cm long Tx line. $x$ denotes the distance of the dielectric from the left end of the Tx line}
			\label{EffetOfDielectric}
	\end{figure}

	 In order to demonstrate the dependence of dielectric constant on resonant frequency, Ethanol ($\epsilon_r\approx25$) \cite{Nia} and Glycerol ($\epsilon_r\approx50$) \cite{Schneider} was used as a dielectric on a 30 cm line. Fig(\ref{EffetOfDielectric}) shows the variations in the resonant frequencies when ethanol and glycerol was used. Table(\ref{resonant freq Compare}) compares the values of the first resonant frequencies as obtained for 3 different dielectrics. The inverse relation between resonant frequency and the relative dielectric constant, as mentioned in section \ref{Impedance Transformation} can be seen. As noted earlier the line is most sensitive to changes done at its center so we observe maximum deviation in frequency when dielectric is placed at 15 cm.
	 
	 \begin{table}[h]
	 	\centering
	 	\begin{tabular}{l l l l}
	 		& \multicolumn{3}{c}{Location of Dielectric}\\
	 		& $x=5cm$ &  $x=10cm$ &  $x=15cm$ \\
	 		\hline\hline
	 		Ethanol & 480MHz & 465MHz & 465MHz \\
	 		Glycerol & 480MHz & 450MHz & 441MHz\\
	 		Water  & 465MHz & 435MHz  & 420MHz\\
	 	\end{tabular}
 	\caption{Measured value of first resonant frequencies at three locations for three dielectrics on a 30cm long Tx line. Here $x$ denotes the distance measured from the left end of Tx line}
 	\label{resonant freq Compare}
	 \end{table}

	 	\paragraph{Discussion for Experiments A and B}It can be deduced from Fig(\ref{S-para Only devices}) that barring the resistive and radiative losses the lines allow the complete transmission of resonant frequencies where as other frequencies are reflected back to the source. The S-parameters of mismatched Tx lines resembles with that of a narrow bandwidth bandpass filter which repeats itself with equal intervals. Changing the dielectric profile of the Tx line helps in shifting the center frequency in a manner which can be closely predicted by the model specified in section \ref{achieving tunability}. This shows the potential of the mismatched line to tune the passband frequencis in a desired way. For a practical application the response of the mismatched line should be subdued at frequencies which are out of the band of interest. This can be achieved by cascading the Tx line with another bandpass filter whose pass band defines the frequencies of interest for a particular application. These two in conjunction can then be used in an application which demands ability to tune the passband along with a narrow bandwidth and a relatively sharp roll-off. 
	 	
	 	 The inherent periodicity of S-Parameters of mismatched Tx lines is seen in linear frequency scale. Thus the slope of the $S_{21}$ is equal for all resonant frequencies when computed in linear scale. Conventionally the roll off is defined as slope of transfer function in logarithmic frequency scale so it is obvious that the roll off calculated from measured values will be steeper at higher frequencies. It was seen that the roll off varies from 100 dB/decade (at lower frequencies) to 280 dB/decade (at higher frequencies).
	 
	 \subsection{Experiment C: Dependence of Resonant Frequency on the Volume of the Dielectric}
	 
\begin{figure}[h!]
	 	\centering
	 	\begin{subfigure}{0.4\textwidth}
	 		\includegraphics[width=6cm]{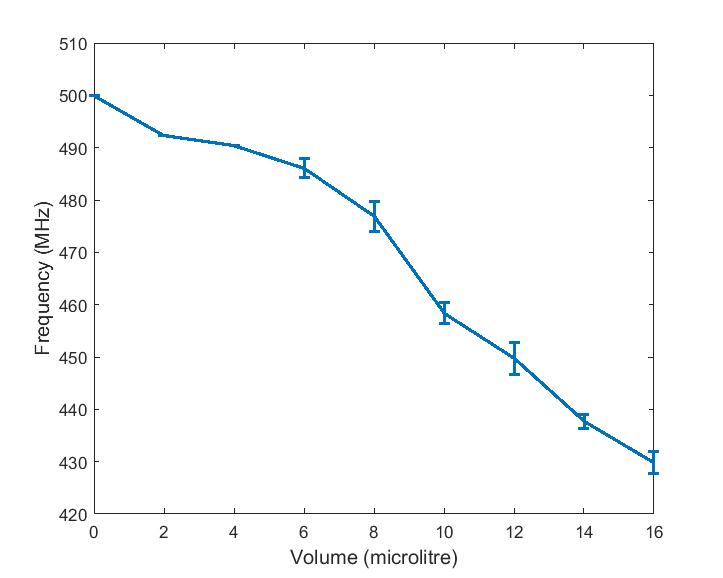}
			\caption{}
			\label{Volume Dependence}
	 	\end{subfigure}
		\begin{subfigure}{0.4\textwidth}
			\includegraphics[width=6cm]{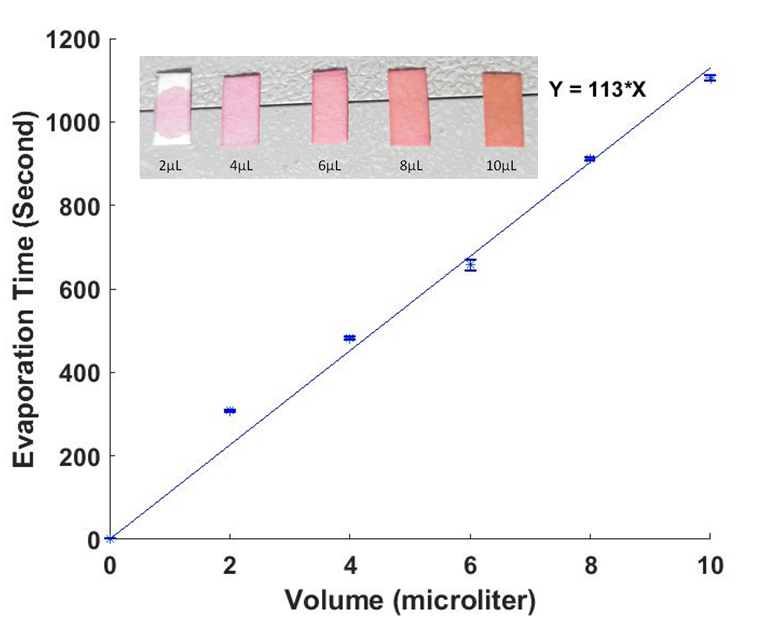}
			\caption{}
			\label{Water Evaporation}
		\end{subfigure}
	\caption{The measurements were performed at 500MHz on a 30cm line. The filter paper of fixed dimensions (10mm $\times$ 5mm) was used in all the measurements. The filter paper was placed at the center to obtain maximum sensitivity. (a) shows the dependence of resonant frequency on volume of water. (b) shows the relation between the time of complete evaporation and the amount of water put onto the paper. A straight line has been fitted ( Rsquare of fit = 0.989) by including the origin of the graph as one of the data points. The images in the inset exemplify the spread of liquid on filter paper for increasing volumes.}
	 \end{figure}

	 The experimental setup adopted to carry out the study enables to see the effect of dielectric volume on the resonant frequencies. The amount of dielectric dispensed on the paper decides effective dielectric constant. This results in variation in the frequency of resonance as one increases the volume of dielectric at \emph{the same location} on the pair of lines.  Fig(\ref{Volume Dependence}) shows the observed changes in the first resonant frequency for a 30cm long Tx line when the amount of water is increased in steps of 2$\mu$L.
	 
	  As dispensed volumes approaches saturation limit, a lump of water is observed due to non-uniform distribution of droplet on filter paper. The lump of water may not remain in exact same position each time the experiment is repeated. The error bars in the results represent the combined effect of non-uniform movement of water on the filter paper each time the experiment is repeated along with the evaporation of liquid from the filter paper surface. To minimize these parameters the experiment was performed in an enclosed environment and the location of disposition of water was controlled to best possible accuracy.  Notice from Fig(\ref{Volume Dependence}) that as the volume is increased the resonant frequency approaches to 420 MHz which is obtained when the paper is completely saturated with water as mentioned in table(\ref{resonant freq Compare}).
	 
	  It is suggestive that the instantaneous volume and its rate of change can be deduced by respectively mapping these to the instantaneous transmitted power and its rate of change at or near a single resonant frequency of an mismatched Tx line. Fig(\ref{Water Evaporation}) demonstrates the case. In this case, after placing the water volume on the paper, resonance frequency shifts to lower value and hence the power received by the load at 500MHz drops. As the water evaporates, the transmission of power rises back to its original value. The time required for its return is plotted against the volume dispensed. An instrument sensitive enough to detect such changes in transmitted power at various frequency can be employed in an application to detect volume variation. The technique has an advantage of being able to measure very low volumes with high accuracy. In addition to this the experiment shows that the phenomenon of evaporation is critical while operating mismatched Tx lines with liquid dielectrics and should be addressed in design process.

	 \section{Conclusion}
	 In the current study it was found that the characteristics behavior of a mismatched Tx line can deliver power to the load at desired resonant frequencies under the controlled conditions. The frequency of resonance depend upon the length of the mismatched line. Further by strategically adjusting the dielectric profile throughout the length of Tx line one can obtain the shift in frequencies of maximum power transmitted. Thus the value of dielectric constant, the length and position of dielectric in the Tx line can provide ability to tune the resonant frequency. To avoid the transmission of power at other resonant frequencies the mismatched line can be cascaded with a bandpass filter of fixed bandwidth. The combination of bandpass filter along with a section of mismatched Tx line can be effectively used as a tunable filter. The technique of using mismatched Tx line shows that it has a potential to be used as a novel sensing mechanism for measuring very small volumes of liquid dielectrics as well as the rate of evaporation of small volumes.
	 
	 \section*{Acknowledgement}
	 Authors of this work would like to thank the Head, Department of Electronics and Instrumentation Science, Savitribai Phule Pune University for providing the necessary resources.

	 \newpage
	 
	 \begin{appendices}
	 	
	 \section{Power Transmission on mismatched Tx line}
	
	 Consider a Tx line with characteristic impedance nearly equal to 347$\Omega$ and real part of propagation constant $\alpha$. It is terminated with a 50$\Omega$ load and it is also driven by a 50$\Omega$ source. 
	 \begin{figure}[h!]
	 	\centering
	 	\includegraphics[scale=0.5]{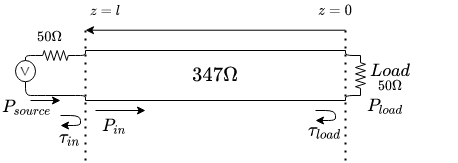}
	 \end{figure}
	 
	 Power delivered to load is \cite{Pozar}
	 \begin{equation}
	 	P_{load}=\frac{1}{2} Re\{V(0)I^*(0)\} = \frac{1}{2} \frac{|V_{0}^+|^2}{Z_0} (1-|\tau_{load}|^2)
	 \end{equation}
 	Power accepted by the Tx line at the generator end $P_{in}$
 	\begin{equation}
 		P_{in} = (1-|\tau_{in}|^2) P_{source}
 	\end{equation}
	 But $P_{in}$ can also be written as 
	 \begin{equation}
	 	P_{in} = \frac{1}{2} Re\{V(z=l)I^*(z=l)\} = \frac{1}{2} \frac{|V_{0}^+|^2}{Z_0} (1-|\tau_{load}|^2) e^{2\alpha l}
 	 \end{equation}
  	
  	Ratio of $P_{load}$ to $P_{in}$ is 
  	\begin{equation}
  		\frac{P_{load}}{P_{in}} = \frac{(1-|\tau_{load}|^2)}{(1-|\tau_{load}|^2) e^{2\alpha l}}
  		\label{PloadtoPin}
  	\end{equation}
  	
  	Thus the ratio of $P_{load}$ to $P_{source}$ is
  	
  	\begin{equation}
  		\frac{P_{load}}{P_{source}} = \frac{(1-|\tau_{load}|^2)}{(1-|\tau_{load}|^2) e^{2\alpha l}} (1-|\tau_{in}|^2)
  		\label{PloadtoPsource}
  	\end{equation}
  	
 \end{appendices}


\end{document}